\documentclass[%
preprint,
superscriptaddress,
nofootinbib,
nobibnotes,
 amsmath,amssymb,
 prl,
]{revtex4-2}

\DeclareUnicodeCharacter{2062}{}
\usepackage{graphicx}% Include figure files
\usepackage{dcolumn}% Align table columns on decimal point
\usepackage{longtable}
\usepackage{bm}% bold math
\usepackage{comment}
\usepackage{lineno}
\usepackage[none]{hyphenat}
\usepackage{hyperref}% add hypertext capabilities
\hypersetup{hidelinks,colorlinks=true,citecolor=blue,linkcolor=blue,breaklinks=true,urlcolor=blue}
\usepackage{xcolor}
\usepackage{amsmath}
\usepackage{float}

\begin{document}

\title{
Superconducting cavity probes sliding ferroelectricity in small-angle twisted WSe$_2$
}

\author{Krishnendu Maji}
\email{kmaji2000@gmail.com}
\affiliation{Department of Condensed Matter Physics and Materials Science, Tata Institute of Fundamental Research, Homi Bhabha Road, Mumbai 400005, India.}
\author{Supriya Mandal}
\affiliation{Université Grenoble Alpes, CNRS, Grenoble INP, Institut Néel, 38000 Grenoble, France}
\author{Sriram H.}
\author{Rishiraj Rajkhowa}
\author{Malhar Date}
\affiliation{Department of Condensed Matter Physics and Materials Science, Tata Institute of Fundamental Research, Homi Bhabha Road, Mumbai 400005, India.}
\author{Sayani Pal}
\affiliation{Department of Physics, Indian Institute of Science, Bangalore 560012, India.}
\author{Meghan P. Patankar}
\affiliation{Department of Condensed Matter Physics and Materials Science, Tata Institute of Fundamental Research, Homi Bhabha Road, Mumbai 400005, India.}
\author{Kenji Watanabe}
\affiliation{Research Center for Functional Materials,
National Institute for Materials Science, 1-1 Namiki, Tsukuba 305-0044, Japan.}
\author{Takashi Taniguchi}
\affiliation{International Center for Materials Nanoarchitectonics, National Institute for Materials Science,  1-1 Namiki, Tsukuba 305-0044, Japan.}
\author{Vibhor Singh}
\affiliation{Department of Physics, Indian Institute of Science, Bangalore 560012, India.}
\author{Mandar M. Deshmukh}
\email{deshmukh@tifr.res.in}
\affiliation{Department of Condensed Matter Physics and Materials Science, Tata Institute of Fundamental Research, Homi Bhabha Road, Mumbai 400005, India.}

\begin{abstract}
\textbf{Ferroelectricity is a property of materials that exhibit 
spontaneous charge polarization. Ferroelectricity in 2D materials is interesting 
because of their applications in memory devices and field-effect transistors. Recently, a new type of ferroelectricity, known as sliding ferroelectricity, has been discovered, in which parallel-stacked bilayers of hexagonal boron nitride (hBN) or transition metal dichalcogenides (TMDCs) develop an out-of-plane electric polarization. In this work, we probe the polarization of small-angle parallel stacked WSe$_2$ by measuring its high-frequency AC response, achieved by embedding it
into a half-wave superconducting coplanar waveguide cavity. We observe a hysteretic response in the capacitance of the stack and quality factor of the cavity, confirming ferroelectric switching in the system. Our results further reveal relaxation effects associated with ferroelectric domain-wall motion. This cavity-based technique has potential applications in probing domain-wall dynamics in a ferroelectric system at high frequencies.}
\end{abstract}

\maketitle

\newpage

Ferroelectricity is a property of certain materials that develop spontaneous electric polarization without requiring an external electric field, and this polarization can be reversed with an external electric field. Ferroelectrics have versatile functional properties, including high dielectric constants, strong piezoelectric effects, and tunable electromechanical responses. These properties make ferroelectrics suitable for a wide range of applications, including non-volatile memory devices \cite{martin_thin-film_2016,guo_non-volatile_2013}, ferroelectric field-effect transistors (FeFETs) \cite{khan_future_2020,jiao_ferroelectric_2023}, and tunable capacitors \cite{he_tunable_2020,mao_ultra-tunable_2024}. Most of the studies on ferroelectricity are based on 3D bulk crystals, especially those with the perovskite structure ABO$_3$, such as barium titanate (BaTiO$_3$) \cite{von_hippel_ferroelectricity_1950} and lead titanate (PbTiO$_3$) \cite{bhide_ferroelectric_1968}. However, 2D materials that host ferroelectricity at room temperature have the advantage over 3D materials for creating miniature and scalable devices for practical applications. Despite those advantages, ferroelectricity in a 2D system poses a unique challenge; the reduced dimensionality enhances the depolarizing fields \cite{junquera_critical_2003,dawber_physics_2005}. However, there are several 2D materials like $\alpha$-In$_2$Se$_3$ \cite{zheng_room_2018} and CuInP$_2$S$_6$ \cite{liu_room-temperature_2016}, which exhibit robust ferroelectric behavior even down to a single layer due to their intrinsic structural asymmetry. Beyond the intrinsically polar monolayers, some van der Waals materials exhibit ferroelectricity in multilayer or heterostructure configurations due to interlayer interactions or stacking-induced symmetry breaking. For instance, materials such as WTe$_2$ \cite{fei_ferroelectric_2018}, and 1T$'$-ReS$_2$ \cite{wan_room-temperature_2022} are centrosymmetric and nonpolar in the monolayer limit, but become ferroelectric when stacked in particular ways.

Recently, a new class of ferroelectricity in 2D materials, called sliding ferroelectricity, has been identified. In this mechanism, ferroelectric polarization arises not from ionic displacements within a monolayer but from the stacking-induced breaking of inversion and out-of-plane mirror symmetries in the system. When two monolayer flakes are stacked in a configuration that breaks both the inversion symmetry and the out-of-plane symmetry of the bilayer, a net out-of-plane polarization emerges due to interfacial charge transfer. Importantly, the relative sliding between the layers does not itself generate ferroelectricity. Instead, sliding switches the system between energetically equivalent stacking configurations with opposite polarization\cite{cao_interlayer_2022,liang_optically_2022}. This form of ferroelectricity was first theoretically proposed \cite{li_binary_2017} and later experimentally demonstrated in systems such as parallel-stacked bilayer hBN \cite{yasuda_stacking-engineered_2021,vizner_stern_interfacial_2021} and various transition metal dichalcogenides (TMDCs) \cite{weston_interfacial_2022,wang_interfacial_2022,deb_cumulative_2022,ko_operando_2023,yang_ferroelectric_2024,liang_shear_2023}. To probe the sliding ferroelectricity in parallel-stacked twisted hBN or TMDC, researchers have used several measurement techniques. Piezoelectric force microscopy (PFM) has been used to visualize interfacial ferroelectric domains and detect switching behavior under an external electric field \cite{wang_interfacial_2022,yasuda_stacking-engineered_2021}. DC transport measurements have also been used to investigate sliding ferroelectricity \cite{wang_interfacial_2022,yasuda_stacking-engineered_2021}. In these experiments, graphene has been employed as a charge sensor, where shifts in its doping reveal the built-in interlayer potential associated with ferroelectric polarization. Capacitance-based sensing offers a complementary approach by directly probing changes in electronic compressibility and polarization through the dielectric response. This method was recently used to investigate spontaneous polarization in bilayer WTe$_2$ as a function of the out-of-plane electric field and carrier density \cite{de_la_barrera_direct_2021}. We next discuss a new device architecture for probing ferroelectric heterostructures.

Superconducting microwave resonators have emerged as powerful probes 
for highly sensitive measurements of the electronic and dielectric 
properties of 2D materials. Their implementations have already been 
used to probe quantum capacitance \cite{maji_superconducting_2024,ranjan_contactless_2017}, 
superfluid stiffness \cite{kreidel_measuring_2024,tanaka_superfluid_2025,banerjee_superfluid_2025}, 
and microwave losses \cite{wang_hexagonal_2022}.
In this work, the cavity-based technique is extended to probe ferroelectricity 
in small-angle (0.1$^\circ$) twisted WSe$_2$ (t-WSe$_2$) using a 
superconducting coplanar waveguide (SCPW) cavity. 
Specifically, a hBN-(t-WSe$_2$)-hBN heterostructure is used as 
one of the coupling capacitors of the cavity. This configuration enables 
us to probe the complex dielectric susceptibility $\tilde{\epsilon}$ of 
the t-WSe$_2$ due to its dispersive and dissipation interaction 
with the microwave cavity mode. 
On one hand, the real part of $\tilde{\epsilon}$ modifies the capacitance
of the heterostructure and thus causes a shift in the resonator frequency. 
On the other hand, the imaginary part of $\tilde{\epsilon}$ modifies the 
losses in the resonator. 
With the change in polarization in t-WSe$_2$ upon application of an 
external out-of-plane electric field, the capacitance of the heterostructure 
changes, which are detected by the change in the resonant frequency 
of the cavity.
Furthermore, as the microwave electric field interacts with the 
ferroelectric domain walls, it induces dissipation by moving these 
domain walls, which is reflected in a reduction of the internal 
quality factor of the cavity.

\begin{figure}[h!]
    \centering
    \includegraphics[width=\textwidth]{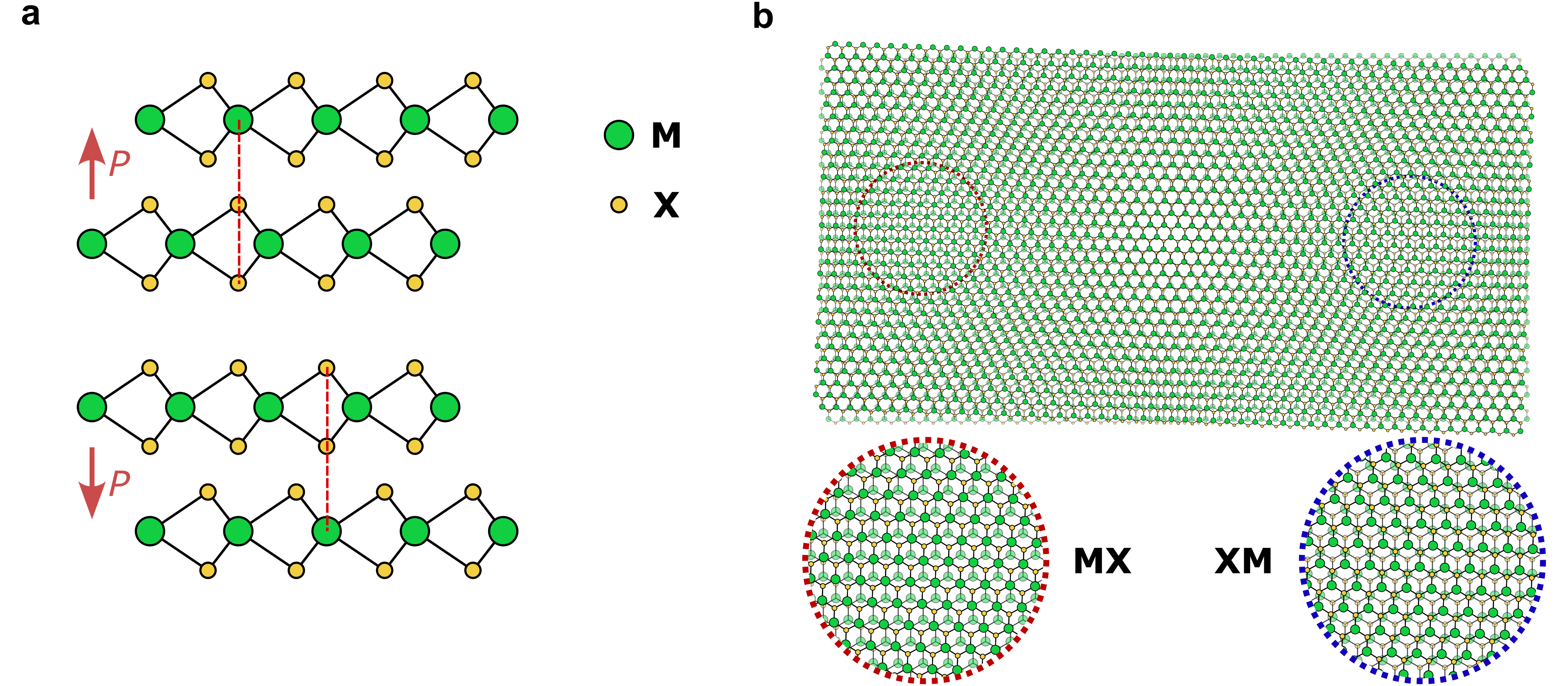}
    \caption[Ferroelectricity in small-angle twisted TMDC]{ \textbf{Ferroelectricity in small-angle twisted TMDC.}  \textbf{a,} Two different stacking orders MX and XM when two monolayer TMDC are stacked in a parallel manner (R-stacked). M and X denote the transition metal and chalcogen atom, respectively. The two stacking orders have opposite polarization ($P$), indicated by the red arrows. \textbf{b,} When two monolayer TMDCs are stacked with a small twisted angle, domains of MX and XM stackings are formed. The two domains are marked by red and blue dashed circles.}
    \label{Figure 1}
\end{figure}

We first discuss basic ideas of inversion symmetry breaking and out-of-plane mirror symmetry breaking that result in interfacial ferroelectricity. Monolayer TMDC is not an inversion-symmetric system. There are two distinct atoms, M and X, in the hexagonal crystal that break the inversion symmetry. However, monolayer TMDCs possess out-of-plane mirror symmetry and, therefore, do not exhibit ferroelectric polarization in the out-of-plane direction. In bilayer TMDC, when two monolayers are stacked in an antiparallel manner (also called hexagonal stacking or H stacking), the inversion symmetry in the system is restored. Consequently, antiparallel stacked (H-stacked) bilayer TMDCs do not show out-of-plane ferroelectricity. In contrast, when two monolayers are stacked in a parallel manner (also called rhombohedral stacking or R stacking), two distinct but energetically equivalent stacking configurations, MX and XM, are possible, as shown in Fig.~\ref{Figure 1}a. In both of these configurations, the system lacks inversion symmetry. Moreover, the out-of-plane mirror symmetry is also broken in these stacking configurations. Because of the lack of inversion symmetry and out-of-plane mirror symmetry, interlayer hybridization occurs between the occupied bands of one layer and the unoccupied bands of the other layer. This leads to charge transfer between the layers \cite{ferreira_weak_2021}, resulting in a net out-of-plane polarization \cite{weston_interfacial_2022,wang_interfacial_2022,yasuda_stacking-engineered_2021,magorrian_multifaceted_2021}. The directions of electric polarization in the MX and XM stackings are opposite to each other, as shown in Fig.~\ref{Figure 1}a.
When two monolayer TMDCs are stacked with a small twist angle, the moiré domains MX and XM are formed, as shown in Fig.~\ref{Figure 1}b. The MX and XM stacking regions are marked by red and blue dashed circles, respectively.
% \section{Experimental methods}
\begin{figure}[h!]
    \centering
    \includegraphics[width=\textwidth]{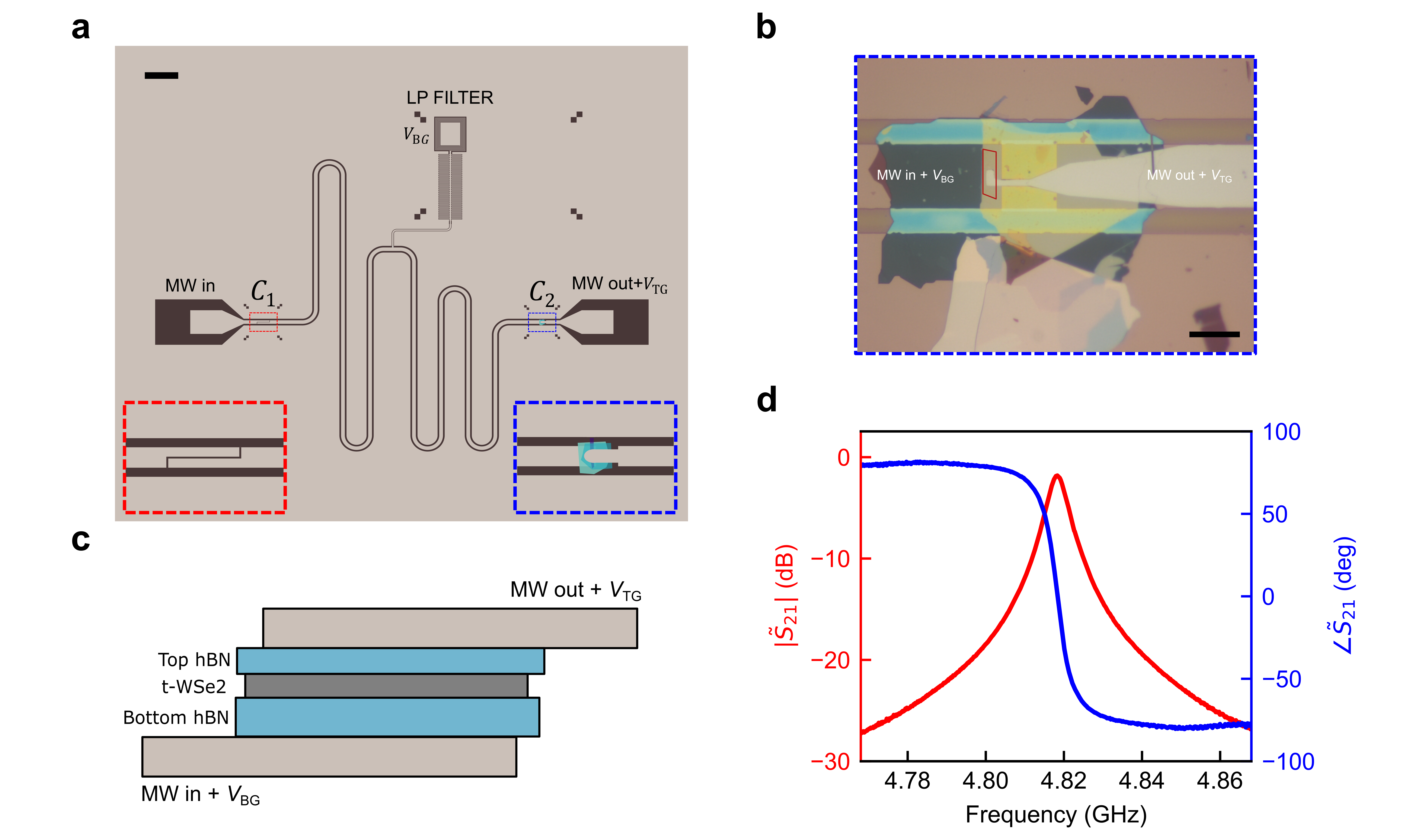}  
   \caption[Integrating the hBN-(t-WSe$_2$)-hBN heterostructure to a half-wave superconducting coplanar waveguide cavity]{\textbf{Integrating the hBN-(t-WSe$_2$)-hBN heterostructure to a half-wave superconducting coplanar waveguide cavity.} 
    \textbf{a,} Representative schematic of the device. The half-wave ($\lambda$/2) coplanar waveguide cavity is designed on a NbTiN-sputtered SiO$_2$ (280~nm)/intrinsic Si (500~$\mathrm{\mu m}$) or sapphire substrate. The inset on the bottom left shows the input coupling capacitor, $C_1$, designed on the left side of the cavity. The inset on the bottom right side shows the zoomed-in image of hBN-(t-WSe$_2$)-hBN heterostructure, which serves as the output coupling capacitor $C_2$. The scale bar on the top left is 200 $\mathrm{\mu m}$.
    \textbf{b,} Optical image of the heterostructure, which is the zoomed-in view of the blue dashed rectangle in panel \textbf{a}. The microwave (MW) signal and applied DC voltage schemes are indicated in the figure. The scale bar is 20~$\mathrm{\mu m}$.
    \textbf{c,} Cross-sectional schematic of the heterostructure. 
    \textbf{d,} Red curve and the blue curve show the magnitude ($|\tilde{S}_{21}|$) and phase ($\angle \tilde{S}_{21}$), respectively, of the transmission coefficient of the cavity when both the gate voltages are set to 0~V. The fundamental resonant frequency is $\sim$4.818~GHz. The cavity shows a $\pi$ phase change at the resonance, which is characteristic of a $\lambda /2$ cavity measured in transmission mode. The data is taken at 1.5~K.}
   \label{Figure 2}
\end{figure}

Fig.~\ref{Figure 2}a shows a representative schematic of the device. The device is fabricated using a 30~nm 
thin film of NbTiN sputtered on SiO$_2$ (280~nm)/intrinsic 
Si (500~$\mathrm{\mu m}$) or sapphire substrates (details of 
the device fabrication are provided in the Supplementary Section~1 
and also in our previous work \cite{maji_superconducting_2024}). 
The input coupling capacitor, $C_1$, is designed to have a 
capacitance of 6.8~fF (as shown in the inset, marked by the 
red dashed box). On the output side, the heterostructure serves 
as the coupling capacitor $C_2$, indicated in the inset 
(blue dashed box). The device is fabricated in a dual-gate configuration. 
The top gate voltage ($V_\mathrm{TG}$) is applied through the microwave 
output port using a bias tee. The bottom gate voltage ($V_\mathrm{BG}$) is 
applied through an on-chip low-pass (LP) filter. The LP filter is 
connected to the central line of the coplanar waveguide at 
the voltage node to minimize microwave loss through it. The details of the LP filter are provided in the Supplementary Section~2.
Fig.~\ref{Figure 2}b shows the 
details of the output coupling capacitor. Fig.~\ref{Figure 2}c shows a schematic of the heterostructure cross-section. The targeted twist-angle between the two WSe$_2$ is $0.1^\circ$. The relative twist angle between the WSe$_2$ and the top hBN in the main device is approximately $7^\circ$, while that between the WSe$_2$ and the bottom hBN is about $6^\circ$. Therefore, WSe$_2$ is well away from the parallel-stacking configuration with both the hBNs.
All measurements are performed in a He-4 cryostat at 1.5~K. 
The details of the measurement setup are given in the Supplementary 
Section~3. To characterize the cavity, we measure the transmission 
$\tilde{S}_{21}$ through the device when both 
gate voltages are set to 0~V. Fig.~\ref{Figure 2}d
shows the magnitude $|\tilde{S}_{21}|$ (in dB unit) and 
phase $\angle\tilde{S}_{21}$ near a resonant mode of the cavity (Supplementary Section~4 
presents details of simulations and the data for a larger 
frequency range that allows us to identify this as the 
fundamental cavity mode).

\begin{figure*}
\centering
\includegraphics[width=140mm]{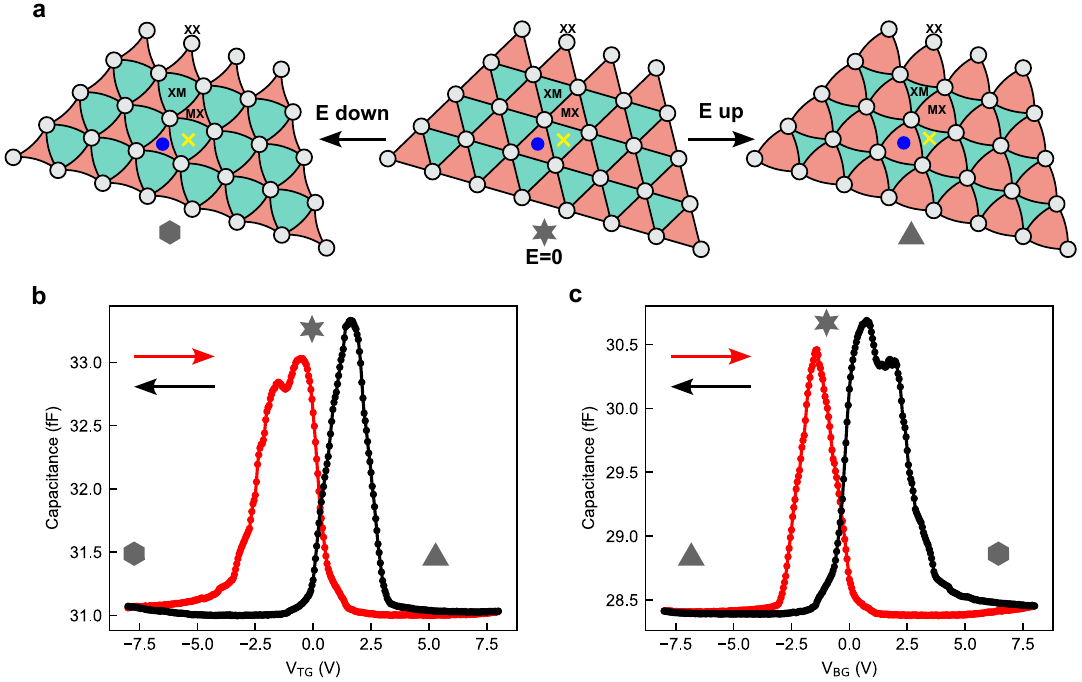}
 \caption[Hysteresis in capacitance with the gate voltages]{\textbf{Hysteresis 
 in capacitance with the gate voltages.} \textbf{a,} Middle panel 
 illustrates the domain structure at zero electric field after relaxation.
 The upward and downward polarizations are shown by blue circles 
 and yellow crosses. The panel on the right/left side shows the 
 shrinking and expansion of the different domains when an upward/downward 
 electric field is applied.
 \textbf{b,} Hysteresis in measured capacitance with the 
 top gate voltage $V_{\mathrm{TG}}$. \textbf{c,} Hysteresis 
 in the measured capacitance with the bottom gate voltage $V_{\mathrm{BG}}$. 
 The data were taken at 1.5~K. The gray star in panel \textbf{a} corresponds 
 to the zero-field condition (middle panel). The gray hexagon and triangle 
 indicate the domain configurations under upward (right panel) and downward 
 (left panel) electric fields, respectively.}
\label{Figure 3}
\end{figure*}

Before the discussion of the experimental data, it is important to highlight
the mechanism responsible for the formation of polarization domains 
in t-WSe$_2$. As noted earlier, Fig.~\ref{Figure 1}b 
shows the domain structure in a t-WSe$_2$ before lattice relaxation. 
The schematic in the middle panel of Fig.~\ref{Figure 3}a 
shows the domain structure of a twisted TMDC with a small twist angle 
in the absence of an external electric field. 
After relaxation, the energetically favorable MX and XM regions grow 
in size, and the unfavorable XX regions shrink in size \cite{magorrian_multifaceted_2021}. 
There are triangular domains of equal area, MX and XM, with opposite 
out-of-plane polarizations, along with smaller XX regions 
at domain intersections. 
The MX and XM domains are separated by domain walls. 
When an external out-of-plane electric field is applied, 
it couples to the layer polarization, energetically favoring 
one stacking order over the other. 
When an upward electric field is applied, it favors the 
MX domains with upward polarization, causing MX domains to expand and 
XM domains to shrink. The reverse effect happens when the direction of
the electric field is reversed (Fig.~\ref{Figure 3}b, left and right panels).

To experimentally probe the system, the transmission $\tilde{S}_{21}$ 
through the cavity for different values of $V_{\mathrm{TG}}$ and 
$V_{\mathrm{BG}}$ is recorded. From the microwave response of the 
fundamental resonator mode, the dispersive and dissipative part, 
namely, the change in capacitance and the internal quality factor 
($Q_i$) of the cavity can be extracted. The response of the cavity can be modeled by the following equation \cite{chen_scattering_2022,pozar_microwave_2012}.
\begin{equation}
    \tilde{S}_{21} = \frac{2}{A + B/Z_0 + CZ_0 + D},
\end{equation}
where $A$, $B$, $C$, and $D$ are the elements of the transmission 
matrix or $ABCD$ matrix. 
The elements of the transmission matrix for a CPW waveguide 
cavity, which consists of a CPW transmission line of length $l$, 
and two coupling capacitors $C_1$ and $C_2$, are given 
by \cite{chen_scattering_2022},
\newline 
$A = \cosh(\gamma l)$ + $\frac{\sinh(\gamma l)}{j\omega C_1 Z_0}$, $B = \sinh(\gamma l)\left(Z_0-\frac{1}{\omega^2 C_1 C_2 Z_0}\right) + \cosh(\gamma l)\left(\frac{1}{j\omega C_1}+\frac{1}{j\omega C_2}\right)$ \newline
$C = \frac{\sinh(\gamma l)}{Z_0}$, $D = \cosh(\gamma l) + \frac{\sinh(\gamma l)}{j\omega C_2 Z
_0}$.
\newline
Here $\gamma = \alpha + j\beta$ is the complex propagation constant, 
and $Z_0$ is the characteristic impedance of the transmission line (matched with the port impedance in this case). 
For a low loss $\lambda/2$ mode, $\beta l$ can be approximated 
as $\pi + \pi (\omega-\omega_0)/\omega_0$ near the resonant frequency, 
where $\omega_0$ is the bare resonant frequency and $\alpha l = \pi/(2Q_i)$, 
where $Q_i$ is the internal quality factor of the cavity \cite{pozar_microwave_2012}. 
The details of the fitting procedure are provided in the Supplementary Section~5.

Next, we investigated the change in capacitance and the quality 
factor as a function of varying gate voltages. Fig.~\ref{Figure 3}b and \ref{Figure 3}c show the measured capacitance with the upward and downward sweep of the top gate voltage and the bottom gate voltage, respectively. There is hysteresis in capacitance with both gate voltages. For both of these measurements, the other gate voltage is fixed at 0~V. The gray star, hexagon, and triangle mark 
the regime corresponding to the zero, upward, and downward electric 
fields. For a small electric field or low gate voltage, a peak in 
capacitance is observed. In the region marked by the grey star, 
there is a peak in capacitance. This corresponds to the situation 
where both MX and XM regions coexist, with opposite out-of-plane 
polarization, as shown in the middle panel of Fig.~\ref{Figure 3}a. It is important to note that in moiré-based ferroelectric systems, complete switching into a monodomain state is not achievable. 
The applied microwave electric field modulates the polarization 
in the system. The domain walls in this region are free to move, 
which increases the effective polarizability of the system. 
For the higher electric field region, marked by the gray hexagon 
and triangle, the MX or XM domain increases in size, depending 
on the direction of the electric field. At a higher electric field, 
most of the domain walls get pinned by defects, impurities, or 
lattice inhomogeneities in the system\cite{weston_interfacial_2022}. 
This results in a decrease in the polarizability of the system, which, 
in turn, reduces the capacitance of the system. Also, it is noticeable 
that there is a symmetry in the two plots. The positive side of the 
top gate voltage qualitatively matches the negative side of the back 
gate voltage and vice versa, implying that the capacitance depends 
on the direction of the electric field.

\begin{figure*}
\centering
\includegraphics[width=120mm]{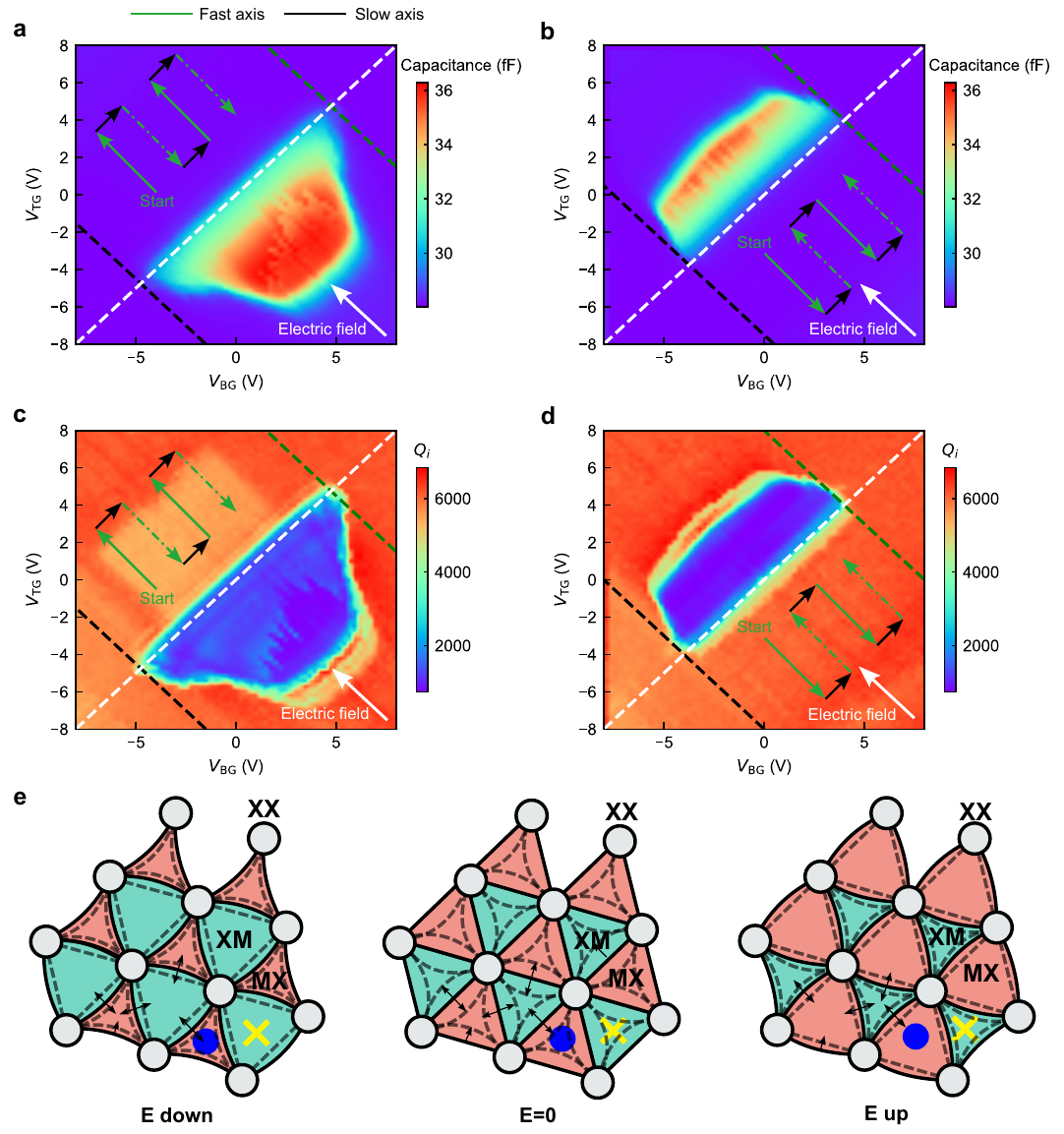}
\caption[Hysteresis with the electric field]{\textbf{Hysteresis with 
the electric field.} (\textbf{a} and \textbf{b}) Color 
plot of capacitance. The solid green arrows indicate the fast axis of the sweep, which in this case lies along the electric field direction. The direction perpendicular to the fast axis is the slow axis of the sweep, which is shown by the solid black arrows. The data shown in the plots were acquired during sweeps along the solid green arrows, and the data acquisition was skipped for sweeps along the dashed-dotted green arrows. A sudden jump in capacitance is observed when the electric field exceeds a critical threshold, accompanied by hysteresis 
between the upward and downward sweeps of the field. 
Similar hysteresis is observed in $Q_i$, as shown in panels 
\textbf{c} and \textbf{d}. The data were taken at 1.5~K. \textbf{e} Schematic of domain wall oscillation induced by the applied 
microwave electric field in different electric field regimes. 
The black dashed curves indicate the oscillation of the domain 
walls from their equilibrium position. At zero electric field, 
the domain walls are free to move, resulting in large oscillations 
of the domain walls from their equilibrium position. 
At a higher electric field, most of the domain walls are pinned, 
resulting in smaller oscillations.}
\label{Figure 4} 
\end{figure*}

Now, the sweep is taken along the electric field axis, as shown in the Fig.~\ref{Figure 4}a and \ref{Figure 4}b. The solid green arrows denote the fast axis of the sweep. The direction perpendicular to the fast axis corresponds to the slow sweep axis, which is denoted by the solid black arrows (The details of the sweep are provided in the Supplementary Section~6). The fast axis lies in the direction of the electric field axis in this case. The data shown in the plots were acquired during sweeps along the solid green arrows, whereas data acquisition was skipped for sweeps along the dashed-dotted green arrows. The data is taken for both directions of the fast sweep within a single scan. The sweeps start from the bottom left (top right) in Fig.~\ref{Figure 4}a (Fig.~\ref{Figure 4}b). Similar notation is used throughout the manuscript. The white dashed line marks the zero electric field. A region of lower capacitance appears in the low electric field region, and hysteresis occurs when the direction of the fast axis is reversed. Additionally, this low capacitance region only appears if the initial electric field is sufficiently large during the sweep. The black and green dashed lines mark the minimum initial electric field required for observing hysteresis. For the upward and downward electric field sweep, the minimum electric fields are $\sim0.16$~V/nm and $\sim0.21$~V/nm, respectively. From the hysteresis loop, we estimate the coercive field to be $\sim0.14$~V/nm.
\begin{table*}[h]
    \centering
    \label{tab:singlecolumn}
    \begin{tabular}{|l|l|c|}
        \hline
        \textbf{Paper} & \textbf{Device} & \textbf{Coercive field} \\
        \hline
        Interfacial ferroelectricity in rhombohedral & Parallel stacked WSe$_2$ & 0.30 V/nm\\
        stacked bilayer transition metal dichalcogenides\cite{wang_interfacial_2022} & & \\
        \hline

        Excitonic signature of ferroelectric order in & Parallel stacked MoS$_2$ & 0.03 V/nm\\
        parallel-stacked MoS$_2$\cite{deb_excitonic_2024} & &\\
        \hline
        Non-volatile electrical electrical polarization & 3R-MoS$_2$ bilayer& 0.65 V/nm\\
        switching via domain wall release in 3R-MoS$_2$ & & \\
        bilayer\cite{yang_non-volatile_2024} & &\\
        \hline
        Our work& Small- angle ($0.1^\circ$) & 0.14 V/nm\\
        & parallel-stacked WSe$_2$ & \\
        \hline
    \end{tabular}
    \caption{Comparison of the coercive fields}
    \label{comparison table}
\end{table*}
Table \ref{comparison table} compares the coercive field obtained in this with values for similar devices reported in literature. Our result is in good agreement with the previously reported values. It is important to note that the coercive field is not an intrinsic property of the material but is highly sample dependent\cite{yang_non-volatile_2024}.
To calculate the electric field, we have used the formula, $\mathrm{electric~field}=\frac{V_{\mathrm{TG}}-V_{\mathrm{BG}}}{t_{\mathrm{thBN}}+t_{\mathrm{bhBN}}}$, where $t_{\mathrm{thBN}}=15$~nm and $t_{\mathrm{bhBN}}=25$~nm are the thicknesses of the top and bottom hBN in our device, respectively. It is to be noted here that the electric field is the macroscopic average electric field as seen by the t-WSe$_2$. The hysteresis loops are notably asymmetric between the positive and negative applied electric fields. In moiré systems, asymmetry between positive and negative applied electric fields is quite common. This asymmetry mainly arises from differences between the top and bottom interfaces, such as the top hBN/WSe$_2$ or hBN/metal interface and the bottom hBN/WSe$_2$ or hBN/metal interface. Although the top and bottom hBN/WSe$_2$ interfaces are expected to be identical, they differ in real devices. In addition, the top and bottom electrodes are fabricated from different superconducting materials with distinct work functions (MoRe for the top electrode and NbTiN for the bottom electrode), which can further contribute to the observed asymmetry \cite{masruroh_asymmetric_2011}. No hysteresis is observed if the sweep is taken with the axis perpendicular to the electric field as the fast axis and axis parallel to the electric field as the slow axis (see Supplementary Section~7). The electric field remains constant during the sweep in both directions along the fast axis. As a result, the domain configuration does not change during the sweep for either direction, and no hysteresis is observed.

The microwave electric field modulates the position of the domain walls from their equilibrium position as indicated by the arrows and black dashed lines in Fig.~\ref{Figure 4}e. This domain wall motion under the microwave electric field is not purely elastic, resulting in dissipation in the system, which changes the internal quality factor of the cavity.

The internal quality factor ($Q_i$) of the cavity is plotted in Fig.~\ref{Figure 4}c and \ref{Figure 4}d. In the region where the capacitance increases, the internal quality factor of the cavity decreases; conversely, the internal quality factor of the cavity increases as the capacitance decreases. At low electric field, the domain walls are free to move as indicated by the large oscillation of the domain walls from their equilibrium position in the middle panel of Fig.~\ref{Figure 4}e. Therefore, the system becomes more lossy at a lower electric field (the blue region in Fig.~\ref{Figure 4}c and \ref{Figure 4}d). At high electric field, most of the domain walls are pinned and respond less to the microwave electric field as indicated by the smaller oscillation of the domain walls in the left and right panels of Fig.~\ref{Figure 4}e.

\begin{figure}[h!]
    \centering
    \includegraphics[width=\textwidth]{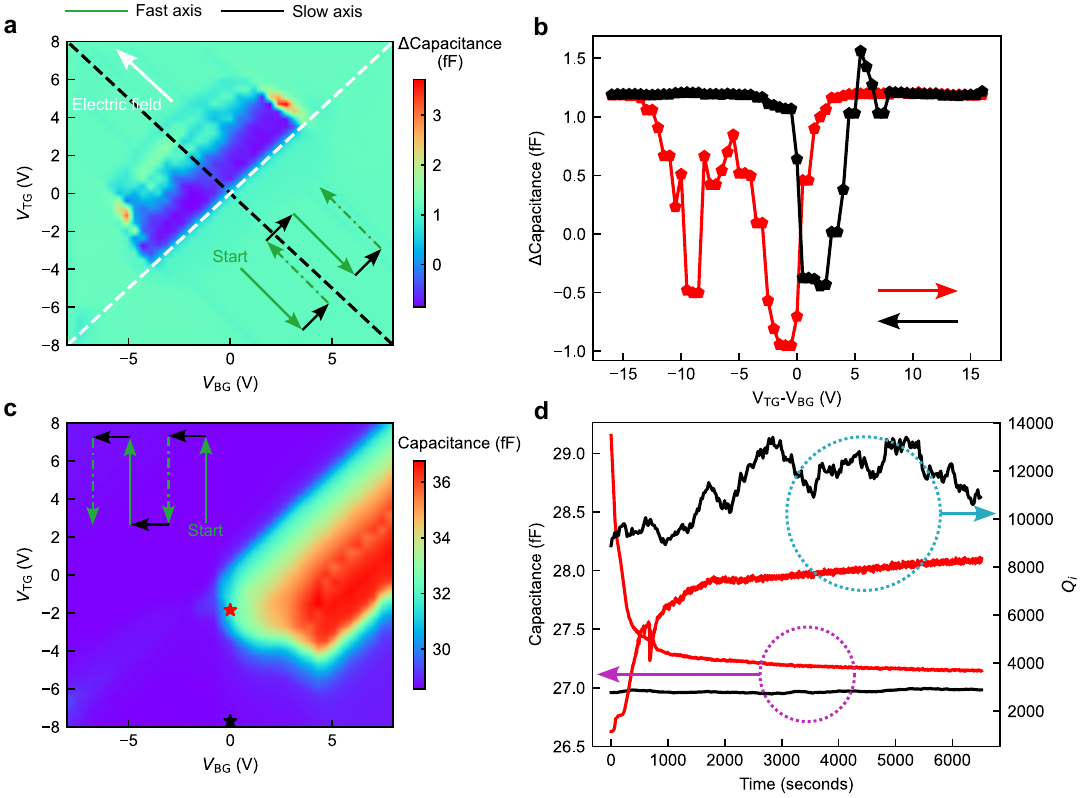}
    \caption[Microwave power dependence and time dependence of the device]{\textbf{Microwave power dependence and time dependence of the device.} \textbf{a,} Difference of the capacitance for downward sweep for microwave powers of -30~dBm and 0~dBm from the VNA. The solid green (black) arrows indicate the fast (slow) sweep axis. Data were acquired along the solid green arrows, and the data acquisition was skipped for sweeps along dashed-dotted green arrows. \textbf{b,} Line cuts for the upward and downward sweep along the black dashed line in panel \textbf{a} with $V_{\mathrm{TG}}-V_{\mathrm{BG}}$, which is proportional to the electric field. \textbf{c,} Capacitance as a function of $V_{\mathrm{TG}}$ and $V_{\mathrm{BG}}$. \textbf{d,} Red (black) curve shows the time dependence of the capacitance and $Q_i$ corresponding to the region marked by the red (black) star in panel \textbf{c}. In the region marked by the red star, the capacitance ($Q_i$) decreases (increases) with time and saturates to a value corresponding to the region marked by the black star. In the region marked by the black star, both the capacitance and $Q_i$ remain almost constant with time. The data were taken at 1.5~K.}
    \label{Figure 5} 
\end{figure}

To investigate the nonlinear response of the domain walls, we performed measurements at different microwave powers. The device is measured at two different powers (-30~dBm and 0~dBm) at the vector network analyzer (VNA) output. Fig.~\ref{Figure 5}a shows the subtracted capacitance data between these two powers for the downward sweep of the electric field. The white dashed line marks the zero electric field. In the higher electric field region, there is a change in the capacitance between the two powers. However, in the low electric field region, the change is less. Fig.~\ref{Figure 5}b presents the line cuts of the capacitance along the black dashed line in Fig.~\ref{Figure 5}a for the upward and downward sweep of the electric field.

Interestingly, the time evolution of the capacitance and $Q_i$ in different regimes shows distinct behaviors. Fig.~\ref{Figure 5}c shows the capacitance data. The device was first biased at the red star point shown in Fig.~\ref{Figure 5}c, and then the gate voltages were kept fixed, and data were taken over time for 1 hour and 50 minutes. The red curves in Fig.~\ref{Figure 5}d show the capacitance and $Q_i$ with time. The capacitance decreases with time and eventually saturates close to a value corresponding to the region marked by the black star in Fig.~\ref{Figure 5}c. Similarly, $Q_i$ increases with time and saturates close to a value corresponding to the region marked by the black star in Fig.~\ref{Figure 5}c. The time evolution of the capacitance and $Q_i$ in the region marked by the black star in Fig.~\ref{Figure 5}c is shown by the black curves in Fig.~\ref{Figure 5}d. Both capacitance and $Q_i$ in this region remain almost constant with time.

The distinct time-evolution behavior of the capacitance and $Q_i$ in the hysteretic low electric-field regime and the non-hysteretic high electric-field regime can be understood as follows. At low electric fields, a mixed-domain state with mobile domain walls exists in the system. When the gate sweep is halted, the previously evolving domain walls gradually relax towards a lower-energy stable configuration by annihilating them. This reduces the effective polarizability of the system, leading to a gradual decrease in capacitance and a simultaneous increase in the $Q_i$. In contrast, at a high electric field, the system is driven into a nearly monodomain state, and most of the domain walls are strongly pinned. Consequently, the capacitance and $Q_i$ remain constant over time.
To confirm that we indeed probed the sliding ferroelectricity at the t-WSe$_2$ interface and not from the hBN or the hBN-hBN interface, we measured a control device (see Supplementary Section~8 for details). In this device, the twisted heterostructure hBN-(t-WSe$_2$)-hBN was replaced with an hBN–hBN heterostructure. No hysteresis is observed in either the capacitance or $Q_i$ with the electric field. The angle of rotation between the two hBN flakes in the second device was set at approximately $23^\circ$, to ensure that the hBN-hBN interface does not produce sliding ferroelectricity. Together, these control experiments demonstrate that the hysteresis observed in our main device originates from the t-WSe$_2$ heterostructure and is not due to the hBN layers or their interface.

To summarize, we demonstrated a high-frequency microwave 
cavity-based technique to probe ferroelectricity in small-angle 
($0.1^\circ$) twisted bilayer WSe$_2$ heterostructures. 
By embedding the hBN-(t-WSe$_2$)-hBN heterostructure as a 
coupling capacitor within a superconducting coplanar waveguide 
cavity, we achieved sensitive measurements of dielectric 
response and energy dissipation at gigahertz frequencies. 
The observed hysteresis in capacitance and internal quality 
factor ($Q_i$) as a function of the applied out-of-plane electric 
field is consistent with ferroelectric switching 
dynamics. In our work, while we did not perform a direction measurement for polarization like PFM, we observed clear and reproducible hysteretic behaviour in the capacitance in the twisted WSe$_2$ devices. Our measurement is sensitive to the net polarization of the system and, consequently, to changes in the relative area fractions of the MX and XM domains. In a real moiré-based sliding ferroelectric system, it is not possible to achieve a fully monodomain state. Therefore, our measurement scheme is well-suited for probing ferroelectricity in such systems. At the same time, we note that parallel stacking alone does not guarantee strong interfacial coupling or the emergence of ferroelectricity, and the observed response may depend on device-specific details. Within this context, our results provide strong and consistent evidence for emergent interfacial ferroelectricity. This microwave cavity-based approach complements 
existing ferroelectric characterization methods such as PFM, 
SHG, and transport measurements. In contrast to these methods, 
our technique can be used to probe the high-frequency dielectric 
properties and loss mechanisms present in a ferroelectric system 
by time-domain measurements.

\section*{acknowledgement}
We thank Vladimir Falko and Moshe Ben Shalom for helpful discussions. We thank Joydip Sarkar and Samrat Ash for experimental assistance. This material is based upon work supported by the Air Force Office of Scientific Research under award number FA2386-25-1-4027. We acknowledge the Nanomission grant SR/NM/NS-45/2016 and DST SUPRA SPR/2019/001247 grant, along with the Department of Atomic Energy of Government of India 12-R\&D-TFR-5.10-0100 for support. The authors also acknowledge the support from J.C. Bose fellowship JCB/2022/000045 from the Department of Science and Technology of India. Preparation of hBN single crystals is supported by the Elemental Strategy Initiative conducted by the MEXT, Japan (Grant Number JPMXP0112101001) and JSPS KAKENHI (Grant Numbers 19H05790 and JP20H00354).

\bibliography{bibliography}

@article{ferreira_weak_2021,
	title = {Weak ferroelectric charge transfer in layer-asymmetric bilayers of {2D} semiconductors},
	volume = {11},
	copyright = {2021 The Author(s)},
	issn = {2045-2322},
	url = {https://www.nature.com/articles/s41598-021-92710-1},
	doi = {10.1038/s41598-021-92710-1},
	
	number = {1},
	urldate = {2024-11-04},
	journal = {Scientific Reports},
	author = {Ferreira, Fábio and Enaldiev, Vladimir V. and Fal’ko, Vladimir I. and Magorrian, Samuel J.},
	month = jun,
	year = {2021},
	pages = {13422},
}

@article{weston_interfacial_2022,
	title = {Interfacial ferroelectricity in marginally twisted {2D} semiconductors},
	volume = {17},
	copyright = {2022 The Author(s)},
	issn = {1748-3395},
	url = {https://www.nature.com/articles/s41565-022-01072-w},
	doi = {10.1038/s41565-022-01072-w},
	
	number = {4},
	urldate = {2024-10-29},
	journal = {Nature Nanotechnology},
	author = {Weston, Astrid and Castanon, Eli G. and Enaldiev, Vladimir and Ferreira, Fábio and Bhattacharjee, Shubhadeep and Xu, Shuigang and Corte-León, Héctor and Wu, Zefei and Clark, Nicholas and Summerfield, Alex and Hashimoto, Teruo and Gao, Yunze and Wang, Wendong and Hamer, Matthew and Read, Harriet and Fumagalli, Laura and Kretinin, Andrey V. and Haigh, Sarah J. and Kazakova, Olga and Geim, A. K. and Fal’ko, Vladimir I. and Gorbachev, Roman},
	month = apr,
	year = {2022},
	pages = {390--395},
}

@article{wang_interfacial_2022,
	title = {Interfacial ferroelectricity in rhombohedral-stacked bilayer transition metal dichalcogenides},
	volume = {17},
	copyright = {2022 The Author(s), under exclusive licence to Springer Nature Limited},
	issn = {1748-3395},
	url = {https://www.nature.com/articles/s41565-021-01059-z},
	doi = {10.1038/s41565-021-01059-z},
	
	number = {4},
	urldate = {2024-10-29},
	journal = {Nature Nanotechnology},
	author = {Wang, Xirui and Yasuda, Kenji and Zhang, Yang and Liu, Song and Watanabe, Kenji and Taniguchi, Takashi and Hone, James and Fu, Liang and Jarillo-Herrero, Pablo},
	month = apr,
	year = {2022},
	pages = {367--371},
}

@article{magorrian_multifaceted_2021,
	title = {Multifaceted moiré superlattice physics in twisted WSe$_2$ bilayers},
	volume = {104},
	url = {https://link.aps.org/doi/10.1103/PhysRevB.104.125440},
	doi = {10.1103/PhysRevB.104.125440},
	number = {12},
	urldate = {2025-04-13},
	journal = {Physical Review B},
	author = {Magorrian, S. J. and Enaldiev, V. V. and Zólyomi, V. and Ferreira, F. and Fal'ko, V. I. and Ruiz-Tijerina, D. A.},
	month = sep,
	year = {2021},
	
	pages = {125440},
}

@article{yasuda_stacking-engineered_2021,
	title = {Stacking-engineered ferroelectricity in bilayer boron nitride},
	volume = {372},
	url = {https://www.science.org/doi/10.1126/science.abd3230},
	doi = {10.1126/science.abd3230},
	number = {6549},
	urldate = {2025-04-07},
	journal = {Science},
	author = {Yasuda, Kenji and Wang, Xirui and Watanabe, Kenji and Taniguchi, Takashi and Jarillo-Herrero, Pablo},
	month = jun,
	year = {2021},
	
	pages = {1458--1462},
}

@article{martin_thin-film_2016,
	title = {Thin-film ferroelectric materials and their applications},
	volume = {2},
	copyright = {2016 Macmillan Publishers Limited},
	issn = {2058-8437},
	url = {https://www.nature.com/articles/natrevmats201687},
	doi = {10.1038/natrevmats.2016.87},
	
	number = {2},
	urldate = {2025-07-20},
	journal = {Nature Reviews Materials},
	author = {Martin, Lane W. and Rappe, Andrew M.},
	month = nov,
	year = {2016},
	pages = {16087},
}

@article{guo_non-volatile_2013,
	title = {Non-volatile memory based on the ferroelectric photovoltaic effect},
	volume = {4},
	copyright = {2013 The Author(s)},
	issn = {2041-1723},
	url = {https://www.nature.com/articles/ncomms2990},
	doi = {10.1038/ncomms2990},
	
	number = {1},
	urldate = {2025-07-20},
	journal = {Nature Communications},
	author = {Guo, Rui and You, Lu and Zhou, Yang and Shiuh Lim, Zhi and Zou, Xi and Chen, Lang and Ramesh, R. and Wang, Junling},
	month = jun,
	year = {2013},
	pages = {1990},
}

@article{khan_future_2020,
	title = {The future of ferroelectric field-effect transistor technology},
	volume = {3},
	copyright = {2020 Springer Nature Limited},
	issn = {2520-1131},
	url = {https://www.nature.com/articles/s41928-020-00492-7},
	doi = {10.1038/s41928-020-00492-7},
	
	number = {10},
	urldate = {2025-07-20},
	journal = {Nature Electronics},
	author = {Khan, Asif Islam and Keshavarzi, Ali and Datta, Suman},
	month = oct,
	year = {2020},
	pages = {588--597},
}

@article{he_tunable_2020,
	title = {A tunable ferroelectric based unreleased {RF} resonator},
	volume = {6},
	copyright = {2020 The Author(s)},
	issn = {2055-7434},
	url = {https://www.nature.com/articles/s41378-019-0110-1},
	doi = {10.1038/s41378-019-0110-1},
	
	number = {1},
	urldate = {2025-07-20},
	journal = {Microsystems \& Nanoengineering},
	author = {He, Yanbo and Bahr, Bichoy and Si, Mengwei and Ye, Peide and Weinstein, Dana},
	month = jan,
	year = {2020},
	pages = {8},
}

@article{mao_ultra-tunable_2024,
	title = {Ultra-tunable dielectric capacitors enhanced by coupling ferroelectric field effect and semiconductor field effect},
	volume = {125},
	issn = {0003-6951},
	url = {https://doi.org/10.1063/5.0227237},
	doi = {10.1063/5.0227237},
	number = {22},
	urldate = {2025-07-20},
	journal = {Applied Physics Letters},
	author = {Mao, Feilong and Gui, Jiashu and Hou, Yongqi and Gao, Siyuan and Zeng, Haohan and Wang, Weibiao and Xu, Zhibin and Zhu, Yifan and Fan, Li and Zhang, Hui},
	month = nov,
	year = {2024},
	pages = {223502},
}

@article{jiao_ferroelectric_2023,
	title = {Ferroelectric field effect transistors for electronics and optoelectronics},
	volume = {10},
	issn = {1931-9401},
	url = {https://doi.org/10.1063/5.0090120},
	doi = {10.1063/5.0090120},
	number = {1},
	urldate = {2025-07-20},
	journal = {Applied Physics Reviews},
	author = {Jiao, Hanxue and Wang, Xudong and Wu, Shuaiqin and Chen, Yan and Chu, Junhao and Wang, Jianlu},
	month = feb,
	year = {2023},
	pages = {011310},
}

@article{von_hippel_ferroelectricity_1950,
	title = {Ferroelectricity, {Domain} {Structure}, and {Phase} {Transitions} of {Barium} {Titanate}},
	volume = {22},
	url = {https://link.aps.org/doi/10.1103/RevModPhys.22.221},
	doi = {10.1103/RevModPhys.22.221},
	number = {3},
	urldate = {2025-07-15},
	journal = {Reviews of Modern Physics},
	author = {von Hippel, A.},
	month = jul,
	year = {1950},

	pages = {221--237},
}

@article{zheng_room_2018,
	title = {Room temperature in-plane ferroelectricity in van der {Waals} {In$_2$Se$_3$}},
	volume = {4},
	url = {https://www.science.org/doi/10.1126/sciadv.aar7720},
	doi = {10.1126/sciadv.aar7720},
	number = {7},
	urldate = {2025-04-20},
	journal = {Science Advances},
	author = {Zheng, Changxi and Yu, Lei and Zhu, Lin and Collins, James L. and Kim, Dohyung and Lou, Yaoding and Xu, Chao and Li, Meng and Wei, Zheng and Zhang, Yupeng and Edmonds, Mark T. and Li, Shiqiang and Seidel, Jan and Zhu, Ye and Liu, Jefferson Zhe and Tang, Wen-Xin and Fuhrer, Michael S.},
	month = jul,
	year = {2018},
	pages = {eaar7720},
}

@article{liu_room-temperature_2016,
	title = {Room-temperature ferroelectricity in {CuInP$_2$S$_6$} ultrathin flakes},
	volume = {7},
	copyright = {2016 The Author(s)},
	issn = {2041-1723},
	url = {https://www.nature.com/articles/ncomms12357},
	doi = {10.1038/ncomms12357},
	
	number = {1},
	urldate = {2025-04-20},
	journal = {Nature Communications},
	author = {Liu, Fucai and You, Lu and Seyler, Kyle L. and Li, Xiaobao and Yu, Peng and Lin, Junhao and Wang, Xuewen and Zhou, Jiadong and Wang, Hong and He, Haiyong and Pantelides, Sokrates T. and Zhou, Wu and Sharma, Pradeep and Xu, Xiaodong and Ajayan, Pulickel M. and Wang, Junling and Liu, Zheng},
	month = aug,
	year = {2016},
	pages = {12357},
}

@article{bhide_ferroelectric_1968,
	title = {Ferroelectric {Properties} of {Lead} {Titanate}},
	volume = {51},
	issn = {1551-2916},
	url = {https://onlinelibrary.wiley.com/doi/abs/10.1111/j.1151-2916.1968.tb13323.x},
	doi = {10.1111/j.1151-2916.1968.tb13323.x},
	
	number = {10},
	urldate = {2025-07-17},
	journal = {Journal of the American Ceramic Society},
	author = {Bhide, V. G. and Hegde, M. S. and Deshmukh, K. G.},
	year = {1968},
	
	pages = {565--568},
}

@article{fei_ferroelectric_2018,
	title = {Ferroelectric switching of a two-dimensional metal},
	volume = {560},
	copyright = {2018 Macmillan Publishers Ltd., part of Springer Nature},
	issn = {1476-4687},
	url = {https://www.nature.com/articles/s41586-018-0336-3},
	doi = {10.1038/s41586-018-0336-3},
	
	number = {7718},
	urldate = {2025-07-15},
	journal = {Nature},
	author = {Fei, Zaiyao and Zhao, Wenjin and Palomaki, Tauno A. and Sun, Bosong and Miller, Moira K. and Zhao, Zhiying and Yan, Jiaqiang and Xu, Xiaodong and Cobden, David H.},
	month = aug,
	year = {2018},
	pages = {336--339},
}

@article{wan_room-temperature_2022,
	title = {Room-Temperature Ferroelectricity in 1⁢T$'$-ReS$_2$ Multilayers},
	volume = {128},
	url = {https://link.aps.org/doi/10.1103/PhysRevLett.128.067601},
	doi = {10.1103/PhysRevLett.128.067601},
	number = {6},
	urldate = {2025-07-15},
	journal = {Physical Review Letters},
	author = {Wan, Yi and Hu, Ting and Mao, Xiaoyu and Fu, Jun and Yuan, Kai and Song, Yu and Gan, Xuetao and Xu, Xiaolong and Xue, Mingzhu and Cheng, Xing and Huang, Chengxi and Yang, Jinbo and Dai, Lun and Zeng, Hualing and Kan, Erjun},
	month = feb,
	year = {2022},

	pages = {067601},
}

@article{junquera_critical_2003,
	title = {Critical thickness for ferroelectricity in perovskite ultrathin films},
	volume = {422},
	copyright = {2003 Macmillan Magazines Ltd.},
	issn = {1476-4687},
	url = {https://www.nature.com/articles/nature01501},
	doi = {10.1038/nature01501},
	
	number = {6931},
	urldate = {2025-07-15},
	journal = {Nature},
	author = {Junquera, Javier and Ghosez, Philippe},
	month = apr,
	year = {2003},
	pages = {506--509},
}

@article{dawber_physics_2005,
	title = {Physics of thin-film ferroelectric oxides},
	volume = {77},
	url = {https://link.aps.org/doi/10.1103/RevModPhys.77.1083},
	doi = {10.1103/RevModPhys.77.1083},
	number = {4},
	urldate = {2025-07-15},
	journal = {Reviews of Modern Physics},
	author = {Dawber, M. and Rabe, K. M. and Scott, J. F.},
	month = oct,
	year = {2005},
	
	pages = {1083--1130},
}

@article{li_binary_2017,
	title = {Binary {Compound} {Bilayer} and {Multilayer} with {Vertical} {Polarizations}: {Two}-{Dimensional} {Ferroelectrics}, {Multiferroics}, and {Nanogenerators}},
	volume = {11},
	issn = {1936-0851},
	shorttitle = {Binary {Compound} {Bilayer} and {Multilayer} with {Vertical} {Polarizations}},
	url = {https://doi.org/10.1021/acsnano.7b02756},
	doi = {10.1021/acsnano.7b02756},
	number = {6},
	urldate = {2025-07-15},
	journal = {ACS Nano},
	author = {Li, Lei and Wu, Menghao},
	month = jun,
	year = {2017},

	pages = {6382--6388},
}

@article{ranjan_contactless_2017,
	title = {Contactless {Microwave} {Characterization} of {Encapsulated} {Graphene} $p$-$n$ {Junctions}},
	volume = {7},
	url = {https://link.aps.org/doi/10.1103/PhysRevApplied.7.054015},
	doi = {10.1103/PhysRevApplied.7.054015},
	number = {5},
	urldate = {2025-05-02},
	journal = {Physical Review Applied},
	author = {Ranjan, V. and Zihlmann, S. and Makk, P. and Watanabe, K. and Taniguchi, T. and Schönenberger, C.},
	month = may,
	year = {2017},
	pages = {054015},
}

@article{tanaka_superfluid_2025,
	title = {Superfluid stiffness of magic-angle twisted bilayer graphene},
	volume = {638},
	copyright = {2025 The Author(s), under exclusive licence to Springer Nature Limited},
	issn = {1476-4687},
	url = {https://www.nature.com/articles/s41586-024-08494-7},
	doi = {10.1038/s41586-024-08494-7},
	number = {8049},
	urldate = {2025-04-06},
	journal = {Nature},
	author = {Tanaka, Miuko and Wang, Joel {\^{I}}-j and Dinh, Thao H. and Rodan-Legrain, Daniel and Zaman, Sameia and Hays, Max and Almanakly, Aziza and Kannan, Bharath and Kim, David K. and Niedzielski, Bethany M. and Serniak, Kyle and Schwartz, Mollie E. and Watanabe, Kenji and Taniguchi, Takashi and Orlando, Terry P. and Gustavsson, Simon and Grover, Jeffrey A. and Jarillo-Herrero, Pablo and Oliver, William D.},
	month = feb,
	year = {2025},
	pages = {99--105},
}

@article{wang_hexagonal_2022,
	title = {Hexagonal boron nitride as a low-loss dielectric for superconducting quantum circuits and qubits},
	volume = {21},
	copyright = {2022 The Author(s), under exclusive licence to Springer Nature Limited},
	issn = {1476-4660},
	url = {https://www.nature.com/articles/s41563-021-01187-w},
	doi = {10.1038/s41563-021-01187-w},
	
	number = {4},
	urldate = {2025-08-26},
	journal = {Nature Materials},
	author = {Wang, Joel I.-J. and Yamoah, Megan A. and Li, Qing and Karamlou, Amir H. and Dinh, Thao and Kannan, Bharath and Braumüller, Jochen and Kim, David and Melville, Alexander J. and Muschinske, Sarah E. and Niedzielski, Bethany M. and Serniak, Kyle and Sung, Youngkyu and Winik, Roni and Yoder, Jonilyn L. and Schwartz, Mollie E. and Watanabe, Kenji and Taniguchi, Takashi and Orlando, Terry P. and Gustavsson, Simon and Jarillo-Herrero, Pablo and Oliver, William D.},
	month = apr,
	year = {2022},
	pages = {398--403},
}

@article{maji_superconducting_2024,
	title = {Superconducting {Cavity}-{Based} {Sensing} of {Band} {Gaps} in {2D} {Materials}},
	volume = {24},
	issn = {1530-6984},
	url = {https://doi.org/10.1021/acs.nanolett.3c04990},
	doi = {10.1021/acs.nanolett.3c04990},
	number = {15},
	urldate = {2025-08-26},
	journal = {Nano Letters},
	author = {Maji, Krishnendu and Sarkar, Joydip and Mandal, Supriya and H., Sriram and Hingankar, Mahesh and Mukherjee, Ayshi and Samal, Soumyajit and Bhattacharjee, Anirban and Patankar, Meghan P. and Watanabe, Kenji and Taniguchi, Takashi and Deshmukh, Mandar M.},
	month = apr,
	year = {2024},
	pages = {4369--4375},
}

@article{vizner_stern_interfacial_2021,
	title = {Interfacial ferroelectricity by van der {Waals} sliding},
	volume = {372},
	url = {https://www.science.org/doi/full/10.1126/science.abe8177},
	doi = {10.1126/science.abe8177},
	number = {6549},
	urldate = {2025-10-06},
	journal = {Science},
	author = {Vizner Stern, M. and Waschitz, Y. and Cao, W. and Nevo, I. and Watanabe, K. and Taniguchi, T. and Sela, E. and Urbakh, M. and Hod, O. and Ben Shalom, M.},
	month = jun,
	year = {2021},
	pages = {1462--1466},
}

@article{kreidel_measuring_2024,
	title = {Measuring kinetic inductance and superfluid stiffness of two-dimensional superconductors using high-quality transmission-line resonators},
	volume = {6},
	url = {https://link.aps.org/doi/10.1103/PhysRevResearch.6.043245},
	doi = {10.1103/PhysRevResearch.6.043245},
	number = {4},
	urldate = {2026-05-11},
	journal = {Physical Review Research},
	author = {Kreidel, Mary and Chu, Xuanjing and Balgley, Jesse and Antony, Abhinandan and Verma, Nishchhal and Ingham, Julian and Ranzani, Leonardo and Queiroz, Raquel and Westervelt, Robert M. and Hone, James and Fong, Kin Chung},
	month = dec,
	year = {2024},
	pages = {043245},
}

@article{banerjee_superfluid_2025,
	title = {Superfluid stiffness of twisted trilayer graphene superconductors},
	volume = {638},
	copyright = {2025 The Author(s), under exclusive licence to Springer Nature Limited},
	issn = {1476-4687},
	url = {https://www.nature.com/articles/s41586-024-08444-3},
	doi = {10.1038/s41586-024-08444-3},
	number = {8049},
	urldate = {2025-10-11},
	journal = {Nature},
	author = {Banerjee, Abhishek and Hao, Zeyu and Kreidel, Mary and Ledwith, Patrick and Phinney, Isabelle and Park, Jeong Min and Zimmerman, Andrew and Wesson, Marie E. and Watanabe, Kenji and Taniguchi, Takashi and Westervelt, Robert M. and Yacoby, Amir and Jarillo-Herrero, Pablo and Volkov, Pavel A. and Vishwanath, Ashvin and Fong, Kin Chung and Kim, Philip},
	month = feb,
	year = {2025},
	pages = {93--98},
}

@book{pozar_microwave_2012,
	title = {Microwave engineering},
	url = {https://search.library.wisc.edu/catalog/9910153599402121},
	publisher = {Fourth edition. Hoboken, NJ : Wiley, [2012] ©2012},
	author = {Pozar, David M},
	year = {2012},
}

@article{chen_scattering_2022,
	title = {Scattering coefficients of superconducting microwave resonators. {I}. {Transfer} matrix approach},
	volume = {106},
	issn = {2469-9950, 2469-9969},
	url = {https://link.aps.org/doi/10.1103/PhysRevB.106.214505},
	doi = {10.1103/PhysRevB.106.214505},
	number = {21},
	urldate = {2025-04-29},
	journal = {Physical Review B},
	author = {Chen, Qi-Ming and Pfeiffer, Meike and Partanen, Matti and Fesquet, Florian and Honasoge, Kedar E. and Kronowetter, Fabian and Nojiri, Yuki and Renger, Michael and Fedorov, Kirill G. and Marx, Achim and Deppe, Frank and Gross, Rudolf},
	month = dec,
	year = {2022},
	pages = {214505},
}

@article{aebischer_inductance_2020,
	title = {Inductance Formula for Rectangular Planar Spiral Inductors with Rectangular Conductor Cross Section},
	volume = {9},
	copyright = {Copyright (c) 2020 H. A. Aebischer},
	issn = {2119-0275},
	url = {https://aemjournals.org/},
	doi = {10.7716/aem.v9i1.1346},
	number = {1},
	urldate = {2025-07-23},
	journal = {Advanced Electromagnetics},
	author = {Aebischer, H. A.},
	month = feb,
	year = {2020},
	pages = {1--18},
}

@article{mi_circuit_2017,
	title = {Circuit quantum electrodynamics architecture for gate-defined quantum dots in silicon},
	volume = {110},
	issn = {0003-6951},
	url = {https://doi.org/10.1063/1.4974536},
	doi = {10.1063/1.4974536},
	number = {4},
	urldate = {2023-11-30},
	journal = {Applied Physics Letters},
	author = {Mi, X. and Cady, J. V. and Zajac, D. M. and Stehlik, J. and Edge, L. F. and Petta, J. R.},
	month = jan,
	year = {2017},
	pages = {043502},
}

@article{goppl_coplanar_2008,
	title = {Coplanar waveguide resonators for circuit quantum electrodynamics},
	volume = {104},
	issn = {0021-8979, 1089-7550},
	url = {http://aip.scitation.org/doi/10.1063/1.3010859},
	doi = {10.1063/1.3010859},
	lp = {en},
	number = {11},
	urldate = {2022-09-12},
	journal = {Journal of Applied Physics},
	author = {Göppl, M. and Fragner, A. and Baur, M. and Bianchetti, R. and Filipp, S. and Fink, J. M. and Leek, P. J. and Puebla, G. and Steffen, L. and Wallraff, A.},
	month = dec,
	year = {2008},
	pages = {113904},
}

@article{deb_excitonic_2024,
	title = {Excitonic signatures of ferroelectric order in parallel-stacked {$\mathrm{MoS_2}$}},
	volume = {15},
	copyright = {2024 The Author(s)},
	issn = {2041-1723},
	url = {https://www.nature.com/articles/s41467-024-52011-3},
	doi = {10.1038/s41467-024-52011-3},
	number = {1},
	urldate = {2025-12-28},
	journal = {Nature Communications},
	author = {Deb, Swarup and Krause, Johannes and Faria Junior, Paulo E. and Kempf, Michael Andreas and Schwartz, Rico and Watanabe, Kenji and Taniguchi, Takashi and Fabian, Jaroslav and Korn, Tobias},
	month = aug,
	year = {2024},
	pages = {7595},
}

@article{yang_non-volatile_2024,
	title = {Non-volatile electrical polarization switching via domain wall release in {3R}-{$\mathrm{MoS_2}$} bilayer},
	volume = {15},
	copyright = {2024 The Author(s)},
	issn = {2041-1723},
	url = {https://www.nature.com/articles/s41467-024-45709-x},
	doi = {10.1038/s41467-024-45709-x},
	number = {1},
	urldate = {2025-12-28},
	journal = {Nature Communications},
	author = {Yang, Dongyang and Liang, Jing and Wu, Jingda and Xiao, Yunhuan and Dadap, Jerry I. and Watanabe, Kenji and Taniguchi, Takashi and Ye, Ziliang},
	month = feb,
	year = {2024},
	pages = {1389},
}

@article{masruroh_asymmetric_2011,
	title = {Asymmetric {Hysteresis} {Loops}, {Leakage} {Current} and {Capacitance} {Voltage} {Behaviors} in {Ferroelectric} {PZT} {Films} {Deposited} on a {Pt}/{$\mathrm{Al_2O_3}$}/{$\mathrm{SiO_2}$}/{Si} {Substrate} by {MOCVD} method with a vapor-deposited {Gold} {Top} {Electrode}},
	issn = {2010362X},
	url = {http://www.ijapm.org/index.php?m=content&c=index&a=show&catid=23&id=248},
	doi = {10.7763/IJAPM.2011.V1.28},
	urldate = {2026-01-08},
	journal = {International Journal of Applied Physics and Mathematics},
	author = {{Masruroh} and Toda, Masayuki},
	year = {2011},
	pages = {144--148},
}

@article{cao_interlayer_2022,
	title = {Interlayer {Registry} {Dictates} {Interfacial} {2D} {Material} {Ferroelectricity}},
	volume = {14},
	issn = {1944-8244},
	url = {https://doi.org/10.1021/acsami.2c20411},
	doi = {10.1021/acsami.2c20411},
	number = {51},
	urldate = {2026-01-13},
	journal = {ACS Applied Materials \& Interfaces},
	author = {Cao, Wei and Hod, Oded and Urbakh, Michael},
	month = dec,
	year = {2022},
	pages = {57492--57499},
}

@article{liang_optically_2022,
	title = {Optically {Probing} the {Asymmetric} {Interlayer} {Coupling} in {Rhombohedral}-{Stacked} $\mathrm{MoS_2}$ {Bilayer}},
	volume = {12},
	url = {https://link.aps.org/doi/10.1103/PhysRevX.12.041005},
	doi = {10.1103/PhysRevX.12.041005},
	number = {4},
	urldate = {2026-01-13},
	journal = {Physical Review X},
	author = {Liang, Jing and Yang, Dongyang and Wu, Jingda and Dadap, Jerry I. and Watanabe, Kenji and Taniguchi, Takashi and Ye, Ziliang},
	month = oct,
	year = {2022},
	pages = {041005},
}

@article{deb_cumulative_2022,
	title = {Cumulative polarization in conductive interfacial ferroelectrics},
	volume = {612},
	copyright = {2022 The Author(s), under exclusive licence to Springer Nature Limited},
	issn = {1476-4687},
	url = {https://www.nature.com/articles/s41586-022-05341-5},
	doi = {10.1038/s41586-022-05341-5},
	number = {7940},
	urldate = {2026-01-13},
	journal = {Nature},
	author = {Deb, Swarup and Cao, Wei and Raab, Noam and Watanabe, Kenji and Taniguchi, Takashi and Goldstein, Moshe and Kronik, Leeor and Urbakh, Michael and Hod, Oded and Ben Shalom, Moshe},
	month = dec,
	year = {2022},
	pages = {465--469},
}

@article{ko_operando_2023,
	title = {Operando electron microscopy investigation of polar domain dynamics in twisted van der {Waals} homobilayers},
	volume = {22},
	copyright = {2023 The Author(s), under exclusive licence to Springer Nature Limited},
	issn = {1476-4660},
	url = {https://www.nature.com/articles/s41563-023-01595-0},
	doi = {10.1038/s41563-023-01595-0},
	number = {8},
	urldate = {2026-01-13},
	journal = {Nature Materials},
	author = {Ko, Kahyun and Yuk, Ayoung and Engelke, Rebecca and Carr, Stephen and Kim, Junhyung and Park, Daesung and Heo, Hoseok and Kim, Hyun-Mi and Kim, Seul-Gi and Kim, Hyeongkeun and Taniguchi, Takashi and Watanabe, Kenji and Park, Hongkun and Kaxiras, Efthimios and Yang, Sang Mo and Kim, Philip and Yoo, Hyobin},
	month = aug,
	year = {2023},
	pages = {992--998},
}

@article{yang_ferroelectric_2024,
	title = {Ferroelectric transistors based on shear-transformation-mediated rhombohedral-stacked molybdenum disulfide},
	volume = {7},
	copyright = {2023 The Author(s), under exclusive licence to Springer Nature Limited},
	issn = {2520-1131},
	url = {https://www.nature.com/articles/s41928-023-01073-0},
	doi = {10.1038/s41928-023-01073-0},
	number = {1},
	urldate = {2026-05-11},
	journal = {Nature Electronics},
	author = {Yang, Tilo H. and Liang, Bor-Wei and Hu, Hsiang-Chi and Chen, Fu-Xiang and Ho, Sheng-Zhu and Chang, Wen-Hao and Yang, Liu and Lo, Han-Chieh and Kuo, Tzu-Hao and Chen, Jyun-Hong and Lin, Po-Yen and Simbulan, Kristan Bryan and Luo, Zhao-Feng and Chang, Alice Chinghsuan and Kuo, Yi-Hao and Ku, Yu-Seng and Chen, Yi-Cheng and Huang, You-Jia and Chang, Yu-Chen and Chiang, Yu-Fan and Lu, Ting-Hua and Lee, Min-Hung and Li, Kai-Shin and Wu, Menghao and Chen, Yi-Chun and Lin, Chun-Liang and Lan, Yann-Wen},
	month = jan,
	year = {2024},
	pages = {29--38},
}

@article{liang_shear_2023,
	title = {Shear {Strain}-{Induced} {Two}-{Dimensional} {Slip} {Avalanches} in {Rhombohedral} {$\mathrm{MoS_2}$}},
	volume = {23},
	issn = {1530-6984},
	url = {https://doi.org/10.1021/acs.nanolett.3c01487},
	doi = {10.1021/acs.nanolett.3c01487},
	number = {15},
	urldate = {2026-01-13},
	journal = {Nano Letters},
	author = {Liang, Jing and Yang, Dongyang and Xiao, Yunhuan and Chen, Sean and Dadap, Jerry I. and Rottler, Joerg and Ye, Ziliang},
	month = aug,
	year = {2023},
	pages = {7228--7235},
}

@article{de_la_barrera_direct_2021,
	title = {Direct measurement of ferroelectric polarization in a tunable semimetal},
	volume = {12},
	copyright = {2021 The Author(s)},
	issn = {2041-1723},
	url = {https://www.nature.com/articles/s41467-021-25587-3},
	doi = {10.1038/s41467-021-25587-3},
	number = {1},
	urldate = {2026-05-11},
	journal = {Nature Communications},
	author = {de la Barrera, Sergio C. and Cao, Qingrui and Gao, Yang and Gao, Yuan and Bheemarasetty, Vineetha S. and Yan, Jiaqiang and Mandrus, David G. and Zhu, Wenguang and Xiao, Di and Hunt, Benjamin M.},
	month = sep,
	year = {2021},
	pages = {5298},
}

\renewcommand{\thefigure}{S\arabic{figure}}
\setcounter{figure}{0}

\newpage
\begin{center}   
{\Huge \textbf{Supplementary Information}}
\end{center}

\section{1. Device Fabrication}
\subsection{Fabrication of the cavity}
The devices were made on intrinsic SiO$_2$ (280~nm)/Si (500~$\mathrm{\mu m}$) or sapphire substrate. The substrate is coated with 30 nm of NbTiN using DC magnetron sputtering. After that, standard photolithography followed by SF$_6$ (12.5~sccm)/Ar (10~sccm) reactive ion etching (RIE) was used to make the meander line. To make the LP filters, standard electron beam lithography followed by RIE was performed.
\subsection{Fabrication of the heterostructures}
To make the heterostructures, first, hBN and WSe$_2$ flakes were exfoliated onto SiO$_2$ (280~nm)/Si (50~$\mu$m) substrates using the mechanical Scotch tape technique. Next, hBN flakes with the desired thickness were selected using optical contrast. To identify monolayer thickness, an optical contrast-based method calibrated with Raman spectroscopy was used. Thereafter, the heterostructures were assembled using a PC (poly(bisphenol A carbonate))/PDMS (polydimethylsiloxane) stamp. Subsequently, the completed stack was placed in the cavity. Finally, the top gate was fabricated using standard electron-beam lithography, followed by MoRe (molybdenum-rhenium) deposition via DC sputtering in a high vacuum chamber at a sputtering pressure of $\sim2\times10^{-3}$~mbar, a bias voltage of 0.45~kV, and a bias current of 0.14~A. With these parameters, the deposition rate for MoRe is 0.5~nm/second.

\section{2. Details of the low-pass filter}
\begin{figure}[H]
    \centering
    \includegraphics[width=\textwidth]{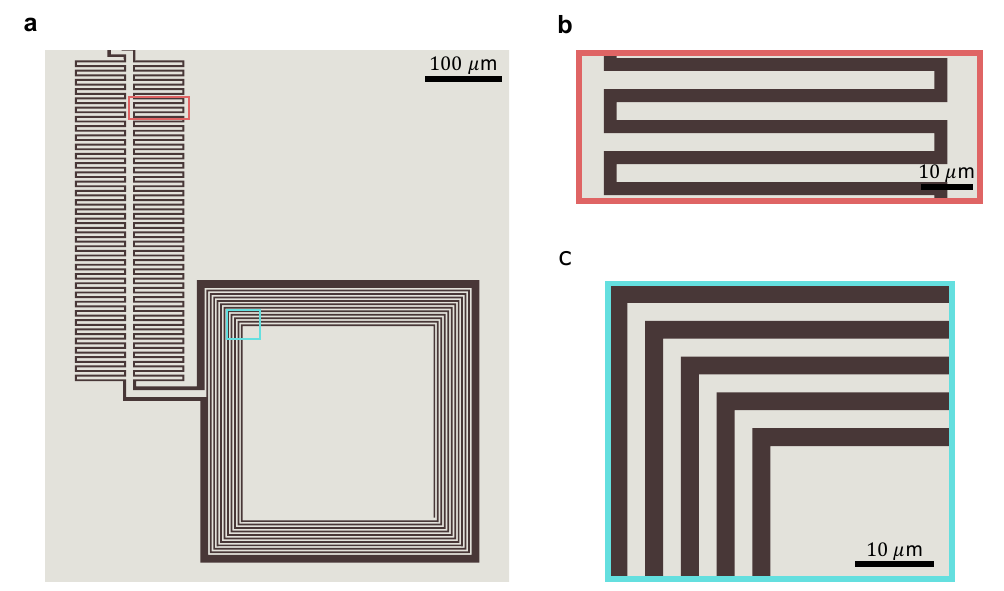}
    \caption[Design of the LP filter]{ \textbf{Design of the LP filter.} \textbf{a,} Shows a schematic of the LP filter, which is an LC circuit composed of an interdigitated capacitor and a spiral inductor. \textbf{b,} Shows the zoomed-in image of the capacitor area marked by a red box in panel \textbf{a}. \textbf{c,} Shows the zoomed-in image of the inductor area marked by the cyan box in panel \textbf{a}.}
    \label{LP filter schematic} 
\end{figure}

We use on-chip low-pass (LP) filters to apply DC voltages to our device. The design of the filter is adopted from the work by Mi et al. \cite{mi_circuit_2017}. The LP filter that we use is an LC filter, as shown in Fig.~\ref{LP filter schematic}a. In our geometry, the capacitor is interdigitated. Fig.~\ref{LP filter schematic}b shows a zoomed-in image of the part of the capacitor marked by the red box in Fig.~\ref{LP filter schematic}a. The inductor is a rectangular spiral inductor. Fig.~\ref{LP filter schematic}c shows a zoomed-in inductor marked by a cyan box in Fig.~\ref{LP filter schematic}a. The inductor is spiral in shape. We have calculated the inductance for our geometry from the formula provided in the paper by Aebischer \cite{aebischer_inductance_2020}. The calculated inductance is 77.63 nH. The capacitance of the interdigitated capacitor can be calculated using the conformal mapping technique \cite{goppl_coplanar_2008}. The calculated value of the capacitance in our geometry is 1.17 pF. The cut-off frequency of the LP filter is $\frac{1}{2\pi\sqrt{LC}}$=490 MHz, which is well below the resonant frequency of the cavity, which is $\sim4.5-5$ GHz.

\begin{figure}[H]
    \centering
    \includegraphics[width=\textwidth]{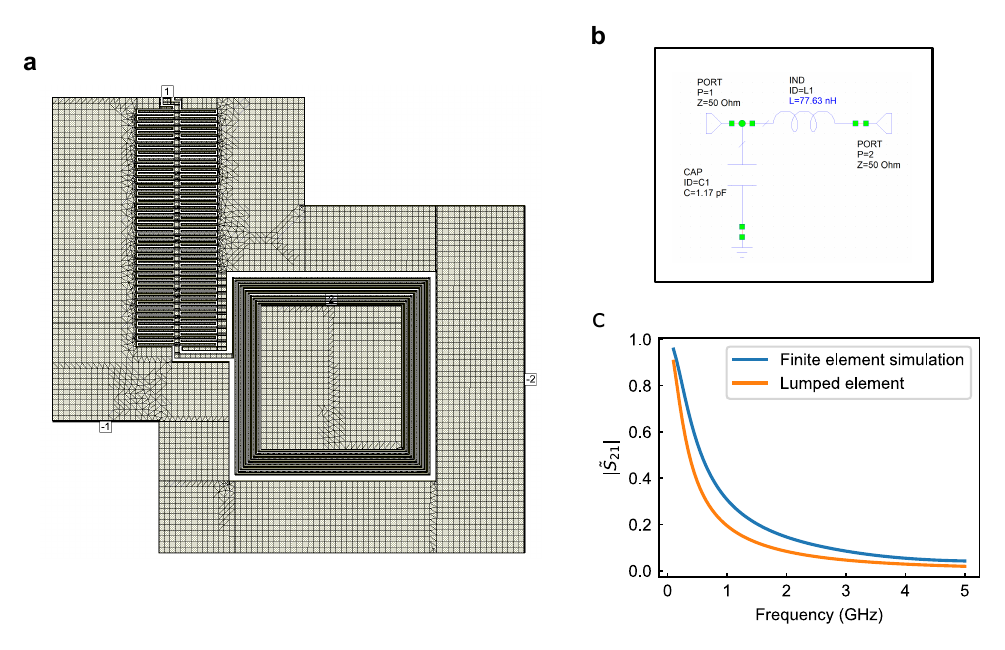}
    \caption[Finite element simulation of the LP filter]{ \textbf{Finite element simulation of the LP filter.} \textbf{a,} Shows the EM structure of the LP filter of the coupling capacitor. \textbf{b,} shows the schematic of the corresponding lumped element circuit model of the LP filter. \textbf{c,}  shows the magnitude ($|\tilde{S}_{21}|$) of the computed $\tilde{S}_{21}$ signal from both the finite element simulation and lumped element circuit.}
    \label{Lp filter simulation} 
\end{figure}

To analyze the performance of the LP filter, we performed a finite element simulation using Cadence Microwave Office. Fig.~\ref{Lp filter simulation}a shows the EM structure of the LP filter for finite element simulation. Fig.~\ref{Lp filter simulation}b shows the schematic of the lumped element circuit corresponding to the LP filter. Fig.~\ref{Lp filter simulation}c shows the magnitude ($|\tilde{S}_{21}|$) of the computed $\tilde{S}_{21}$ signal from the finite element simulation and lumped element circuit. Although the qualitative features of both traces look similar, they do not overlap perfectly. This discrepancy arises because the formulas used to calculate the capacitance and inductance are approximate, and the finite element simulation also has its limitations.

\section{3. Measurement Details}
\begin{figure}[H]
    \centering
    \includegraphics[width=\textwidth]{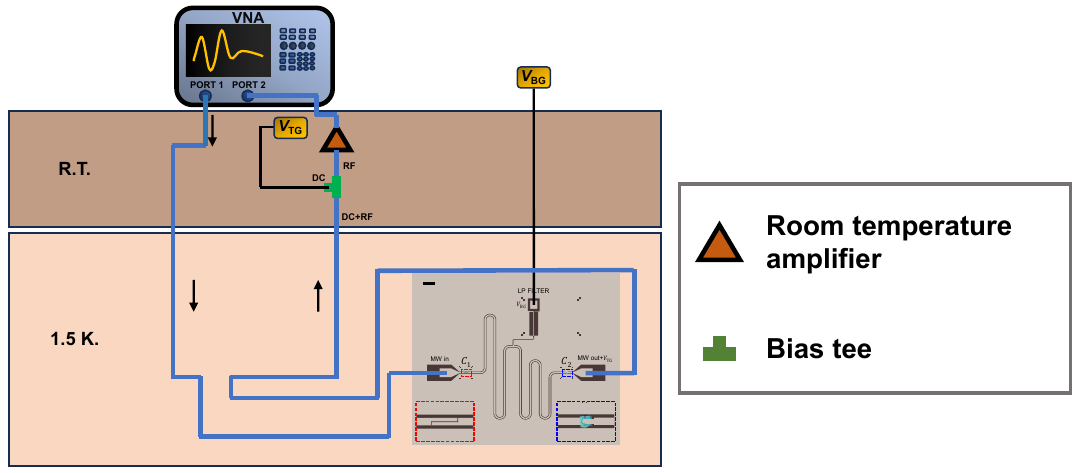}
   \caption[Microwave measurement setup for measurements in the He-4 wet cryostat]{ \textbf{Microwave measurement setup for measurements in the He-4 wet cryostat.} shows the measurement setup for measurement in the He-4 wet cryostat. The base temperature is 1.5~K. The output microwave signal is amplified at room temperature using a 25~dB Mini-Circuits amplifier (ZVA-213-S+). The top gate voltage is supplied through the output port using a bias tee.}
   \label{Supp cryostat wiring}
\end{figure}

All measurements were performed in a He-4 cryostat at a base temperature of 1.5~K, with the measurement circuit shown in Fig.~\ref{Supp cryostat wiring}. A vector network analyzer (R\&S ZNB 20) measured the transmission coefficient, $\tilde{S}_{21}$, of the device. The microwave output signal was amplified at room temperature using a 25~dB Mini-Circuits amplifier (ZVA-213-S+). DC gate voltages were applied via a National Instruments DAQ. The back gate voltage was supplied from a DC line, while the top gate voltage was supplied through the output port using a bias tee.

\section{4. Microwave simulation and cavity characteristics}
\begin{figure}[h!]
    \centering
    \includegraphics[width=0.7\textwidth]{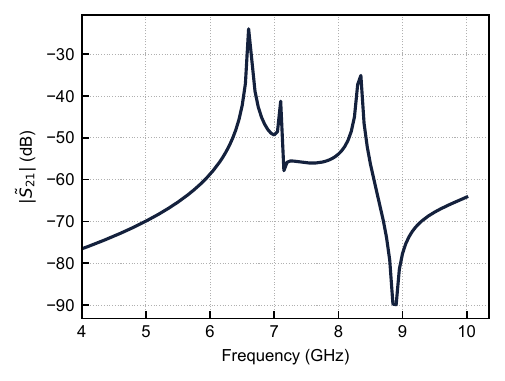}
    \caption[Simulation data of the cavity]{\textbf{Simulation data of the cavity.} Show the $|\tilde{S}_{21}|$ data obtained from the microwave simulation using the AWR Microwave Office software.}
    \label{supp simulation}
\end{figure}

\begin{figure}[h!]
    \centering
    \includegraphics[width=\textwidth]{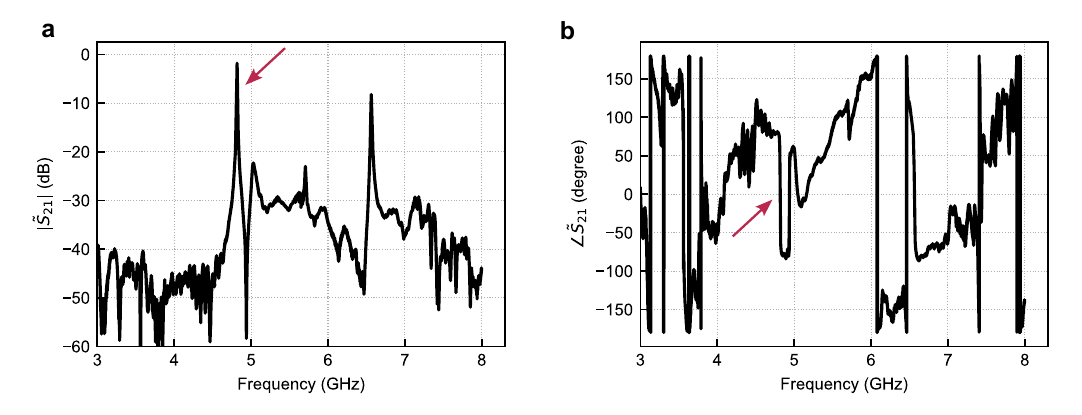}
    \caption[Frequency spectrum of the cavity]{\textbf{Frequency spectrum of the cavity.} \textbf{a} and \textbf{b} show the magnitude $|\tilde{S}_{21}|$ (in dB unit) and phase $\angle \tilde{S}_{21}$ of the $\tilde{S}_{21}$ data, respectively, over a wide range of frequency. The red arrows annotate the fundamental mode of the cavity.}
    \label{supp cavity} 
\end{figure}

We have simulated the frequency response of the cavity using AWR Microwave Office software. Fig.~\ref{supp simulation} shows the simulated frequency response of the cavity with filters in place and with an identical coupling capacitor of 6.8~fF on both the input and output sides. Two peaks are visible, corresponding to different cavity modes. The first peak reaches the maximum transmission and is therefore identified as the fundamental mode of the cavity.
Fig.~\ref{supp cavity}a and \ref{supp cavity}b show the magnitude $|\tilde{S}_{21}|$ (in dB unit) and phase $\angle \tilde{S}_{21}$ of the experimental $\tilde{S}_{21}$ data for the cavity coupled to the hBN-(t-WSe$_2$)-hBN heterostructure, respectively, with both the gate voltages fixed to 0~V. The $|\tilde{S}_{21}|$ data qualitatively match well with the simulation. During the measurements, we focused on the peak annotated by the red arrow, corresponding to the fundamental mode of the cavity. In Fig.~\ref{supp cavity}b a 180-degree phase shift is observed around the peak, which is characteristic of a half-wave ($\lambda/2$) cavity. 

\section{5. Fitting Procedure}
The data were fitted using the following equation
\begin{equation}
    \tilde {S}_{21} = \frac{2}{A + B/Z_0 + CZ_0 + D},
    \label{Supp eq 1}
\end{equation}
where $A$, $B$, $C$, and $D$ are the elements of the transmission matrix or $ABCD$ matrix \newline 
$A = \cosh(\gamma l)$ + $\frac{\sinh(\gamma l)}{j\omega C_1 Z_0}$,~~~~~$B = \sinh(\gamma l)\left(Z_0-\frac{1}{\omega^2 C_1 C_2 Z_0}\right) + \cosh(\gamma l)\left(\frac{1}{j\omega C_1}+\frac{1}{j\omega C_2}\right)$ \newline
$C = \frac{\sinh(\gamma l)}{Z_0}$,~~~~~$D = \cosh(\gamma l) + \frac{\sinh(\gamma l)}{j\omega C_2 Z
_0}$
\newline
where $l$ is the length of the transmission line, $\gamma = \alpha + j\beta$ is the complex propagation constant of the microwave field, $Z_0$ is the characteristic impedance of the transmission line. For $\lambda/2$ cavity $\beta l = \pi + \pi (\omega-\omega_0)/\omega_0$, where $\omega_0$ is the bare resonant frequency and $\alpha l = \pi/(2Q_i)$, where $Q_i$ is the internal quality factor of the cavity. We multiply Eq. \ref{Supp eq 1} with a factor $Ae^{-j \omega \tau}$, which accounts for the cable loss and finite length of the feedlines.

\begin{figure}[h!]
    \centering
    \includegraphics[width=\textwidth]{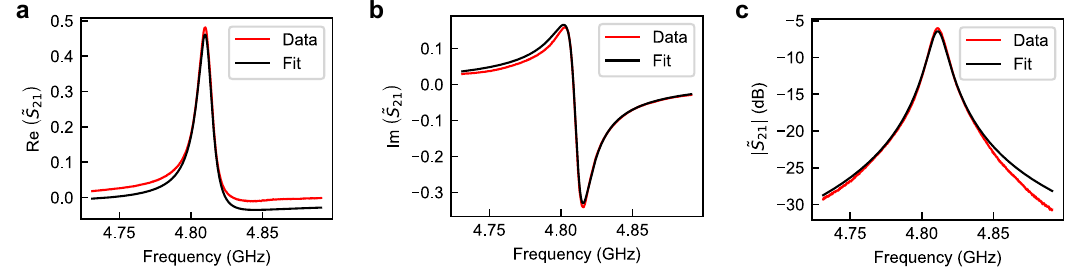}
    \caption[Fitting of the $\tilde{S}_{21}$ data]{\textbf{Fitting of the $\tilde{S}_{21}$ data.} \textbf{a},\textbf{b}, and \textbf{c} show the real part, imaginary part, and magnitude (in dB unit) of the $\tilde{S}_{21}$ data at $V_{\mathrm{TG}}=V_{\mathrm{BG}}=0$, along with their corresponding fits.}
    \label{fitting}
\end{figure}

For fitting, both the real and imaginary components of the measured $\tilde{S}_{21}$ data were used. $C_2$, $Q_i$, $\tau$, and $\omega_0$ were taken as fitting parameters. For each $(V_\mathrm{TG}, ~V_\mathrm{BG})$, the $\tilde{S}_{21}$ data were fitted using $C_2$, $Q_i$, and $\tau$ as fitting parameters, while keeping the bare resonant frequency $\omega_0$ fixed. This ensures that any observed shift in the resonant frequency arises solely from the capacitive loading due to $C_2$, and not from variations in $\omega_0$. This fitting procedure allows us to extract both the effective capacitance $C_2$ of the heterostructure and the internal quality factor $Q_i$, which reflects the loss mechanisms in the system. Fig.~\ref{fitting}a, \ref{fitting}b, and \ref{fitting}c show the real part, imaginary part, and magnitude (in dB unit) of $\tilde{S}_{21}$ data, respectively, along with their corresponding fits for a reference point (0,0) in ($V_{\mathrm{TG}}$, $V_{\mathrm{BG}}$) space.
The extracted capacitance and internal quality factor from the fitting are $C_2=37.207$ fF and $Q_i=1736.092$. The errors associated with them are $\delta C_2=14$ aF and $\delta Q_i=32.710$. Therefore, the fitting uncertainties of the extracted capacitance and internal quality factor from the fitting are of the order of magnitude lower than the extracted values.

\section{6. Details of the sweep}
The 2D data shown in Fig.~4a-4d, 5a, and 5c in the main manuscript and Fig. \ref{Sup fig perp electric field}, \ref{Sup fig hBN-hBN}, and \ref{Sup fig 1L-Wse2} in this Supplementary Information were taken in the following way.

\begin{enumerate}
\item The gate voltage is swept along the fast axis at a ramp rate of 30~V/min with a bin size of 0.05~V. At each voltage point, the complex transmission parameter $\tilde{S}_{21}$ is measured as a function of frequency.
\item The sweep direction along the fast axis is then reversed, and data are acquired in the opposite direction using the same procedure described in step~1.
\item After completing both forward and backward sweeps along the fast axis, the gate voltage is incremented along the slow axis at a ramp rate of 30~V/min with a bin size of 0.05~V.
\item Steps~1–3 are repeated until the full range along the slow axis is covered.
\end{enumerate}

Acquiring a single frequency spectrum takes $\sim1.21$~s. From this, we estimate the effective sweep rates along the two axes. The sweep rate along the fast axis is $\sim2.5$~V/min, whereas the sweep rate along the slow axis is $\sim0.01$~V/min.

\section{7. Data for fast sweep along the constant electric field}
Unless stated otherwise, solid green (black) arrows indicate the fast (slow) sweep axis. Data were acquired along the solid green arrows, and data acquisition was skipped along the dashed-dotted green arrows. All measurements are performed at 1.5~K.

\begin{figure}[H]
    \centering
    \includegraphics[width=\textwidth]{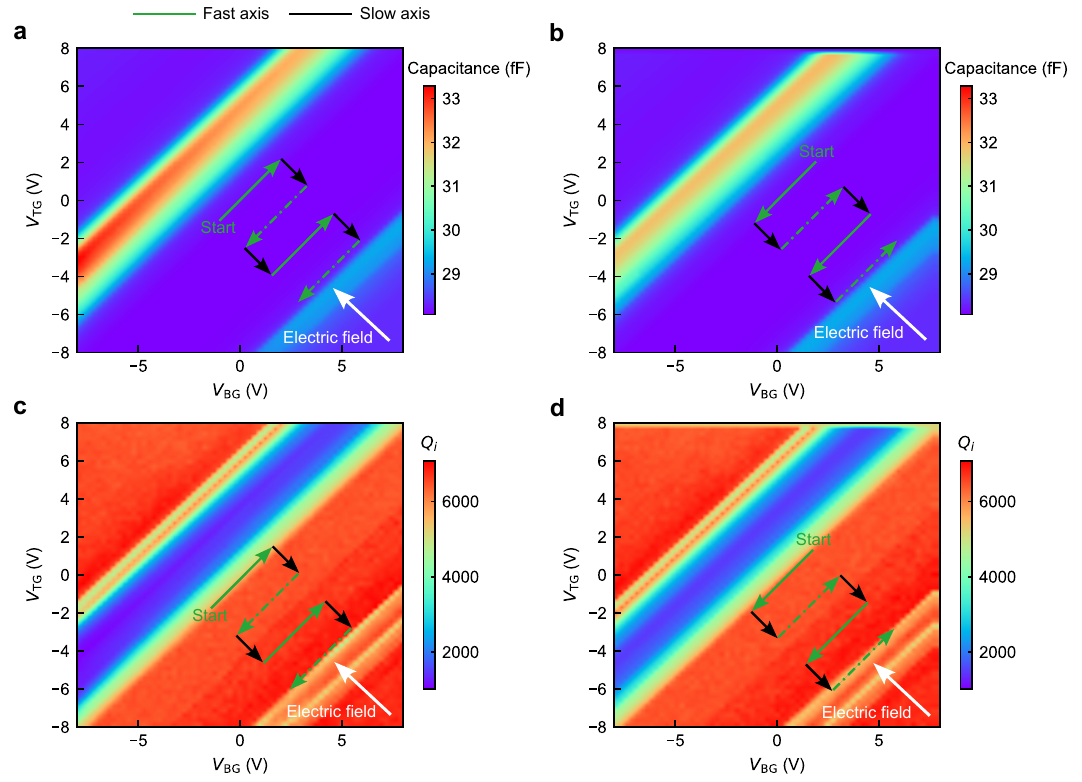}
    \caption[Data for fast sweeps along the constant electric field]{\textbf{Data for fast sweeps along the constant electric field.} \textbf{a} and \textbf{b} show the color plot of capacitance. There is no hysteresis in the capacitance between the two different fast axes. \textbf{c} and \textbf{d} show the $Q_i$ corresponding to \textbf{a} and \textbf{b}, respectively. There is no hysteresis in the $Q_i$ also between the two fast axes. The data were taken at 1.5~K.}
    \label{Sup fig perp electric field}
\end{figure}

Fig.~\ref{Sup fig perp electric field} shows the data with the fast axis perpendicular to the electric field and the slow axis parallel to the electric field. No hysteresis is observed in capacitance or in $Q_i$ between two sweep directions. The electric field remains constant during the sweep in both directions along the fast axis. As a result, the domain configuration does not change during the sweep for either direction, and no hysteresis is observed.

\section{8. Control Devices}
To confirm that hysteresis originates from twisted WSe$_2$, a control device is measured.
\subsection{hBN-hBN heterostructure}
\begin{figure}[H]
    \centering
    \includegraphics[width=\textwidth]{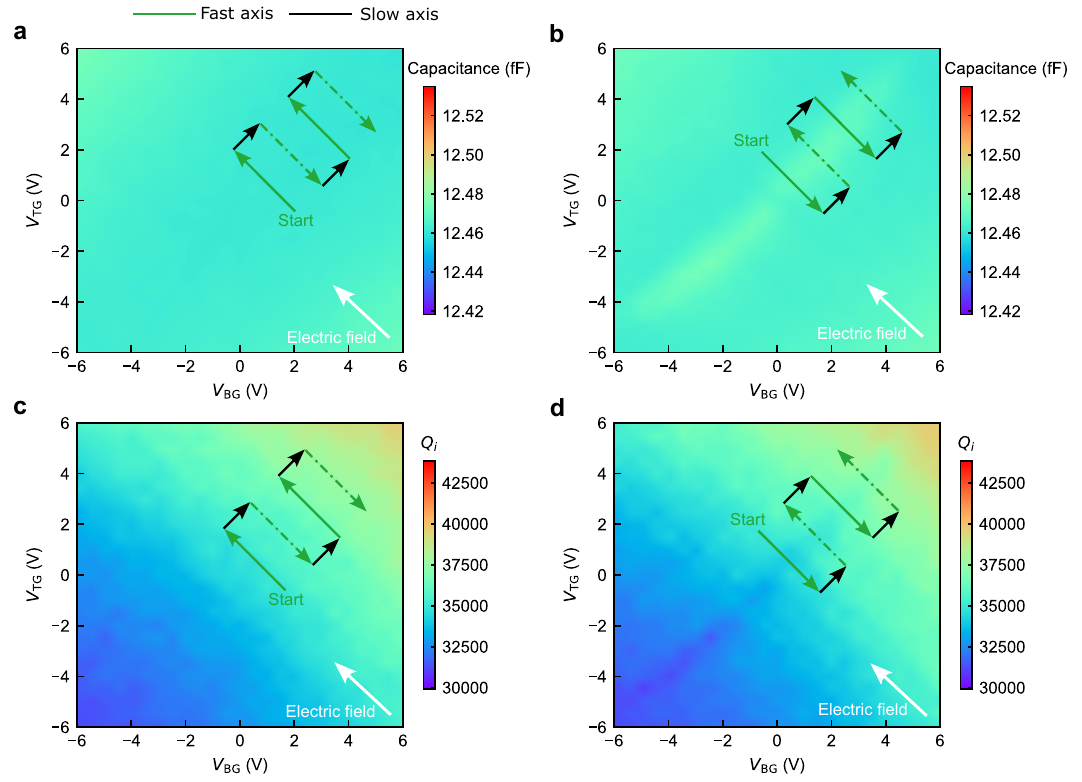}
    \caption[Absence of hysteresis in the control hBN-hBN device]{\textbf{Absence of hysteresis in the control hBN-hBN device.} \textbf{a} and \textbf{b} show the color plot of capacitance. No hysteresis is observed along the electric field. \textbf{c} and \textbf{d} show the $Q_i$ corresponding to \textbf{a} and \textbf{b} respectively. No hysteresis is observed in $Q_i$ also. The data were taken at 1.5~K.}
    \label{Sup fig hBN-hBN}
\end{figure}

To confirm that hysteresis does not originate from the hBN-hBN interface or hBN, a control device was measured where the hBN-(t-WSe$_2$)-hBN heterostructure is replaced by an hBN-hBN heterostructure. The targeted twist angle between the two hBN flakes was $\sim 23^\circ$ to ensure that ferroelectricity does not arise from the hBN-hBN interface due to sliding ferroelectricity\cite{yasuda_stacking-engineered_2021}. Fig.~\ref{Sup fig hBN-hBN}a and \ref{Sup fig hBN-hBN}b show the measured capacitance for this device. There is no hysteresis in the capacitance. The corresponding internal quality factor measurements $Q_i$, shown in Fig.~\ref{Sup fig hBN-hBN}c and \ref{Sup fig hBN-hBN}d, also do not show hysteresis, confirming that the hBN–hBN interface or hBN does not contribute to the observed ferroelectric response.
\subsection{Monolayer WSe$_2$}
\begin{figure}[H]
    \centering
    \includegraphics[width=\textwidth]{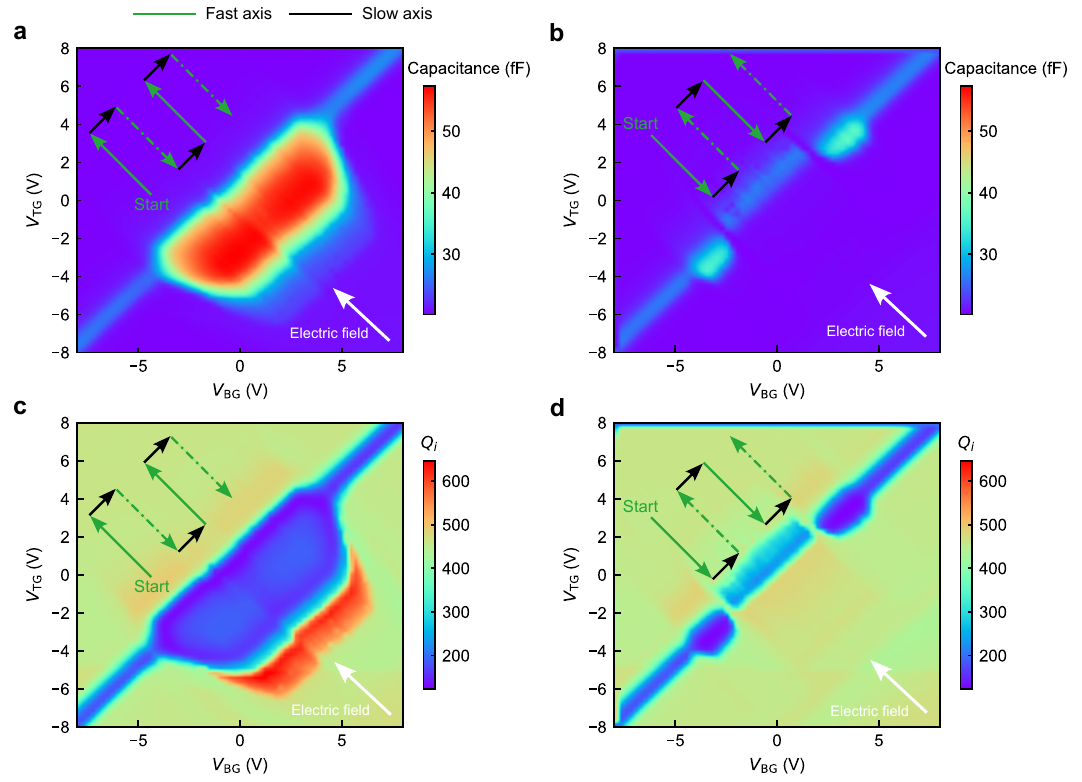}
    \caption[Hysteresis in the hBN-1L WSe$_2$-hBN device.]{\textbf{Hysteresis in the hBN-1L WSe$_2$-hBN device.} \textbf{a} and \textbf{b} show the color plot of capacitance. \textbf{c} and \textbf{d} show the $Q_i$ corresponding to \textbf{a} and \textbf{b}, respectively. A similar hysteresis is observed in $Q_i$ also. The data were taken at 1.5~K.}
    \label{Sup fig 1L-Wse2}
\end{figure}

We also measured a device in which the twisted WSe$_2$ (t-WSe$_2$) is replaced by a monolayer WSe$_2$ (1L-WSe$_2$). Fig~\ref{Sup fig 1L-Wse2}a and \ref{Sup fig 1L-Wse2}b show the capacitance map. Fig.~\ref{Sup fig 1L-Wse2}c and \ref{Sup fig 1L-Wse2}d present the $Q_i$ map.

\end{document}